\documentclass[preprint,amsmath,amssymb,aps,prd,nofootinbib]{revtex4}
\usepackage{epsfig,graphicx,color,appendix}
\begin{document}
%\begin{flushright}
%IBS-CTPU-18-
%\end{flushright}
\def\CP{{\it CP}~}
\def\cp{{\it CP}}
\title{\mbox{}\\[10pt]
Fermion masses and flavor mixings\\ and strong CP problem}

\author{Y. H. Ahn}
\affiliation{Key Laboratory of Particle Astrophysics, Institute of High Energy Physics,
Chinese Academy of Sciences, Beijing, 100049, China}
\email{axionahn@naver.com}

%\date{\today}% It is always \today, today,
             %  but any date may be explicitly specified

\begin{abstract}
\noindent  For all the success of the Standard Model (SM), it is on the verge of being surpassed. In this regard we argue, by showing a minimal flavor-structured model based on the non-Abelian discrete $SL_2(F_3)$ symmetry, that $U(1)$ mixed-gravitational anomaly cancellation could be of central importance in constraining the fermion contents of a new chiral gauge theory. % realized in gauge/gravity duality. 
Such anomaly-free condition together with the SM flavor structure demands a condition $k_1\,X_1/2=k_2\,X_2$ with $X_i$ being a charge of $U(1)_{X_i}$ and $k_i$ being an integer, both of which are flavor dependent. We show that axionic domain-wall condition $N_{\rm DW}$ with the anomaly free-condition depends on both $U(1)_X$ charged quark and lepton flavors; the seesaw scale congruent to the scale of Peccei-Quinn symmetry breakdown can be constrained through constraints coming from astrophysics and particle physics.  
Then the model extended by $SL_2(F_3)\times U(1)_X$ symmetry can well be flavor-structured in a unique way that $N_{\rm DW}=1$ with the $U(1)_X$ mixed-gravitational anomaly-free condition demands additional Majorana fermion and the flavor puzzles of SM are well delineated by new expansion parameters expressed in terms of $U(1)_X$ charges and $U(1)_X$-$[SU(3)_C]^2$ anomaly coefficients. And the model provides remarkable results on neutrino (hierarchical mass spectra and unmeasurable neutrinoless-double-beta decay rate together with the predictions on atmospheric mixing angle and leptonic Dirac CP phase favored by the recent long-baseline neutrino experiments), QCD axion, and flavored-axion. 

\end{abstract}
\maketitle %
%%%%%%%%%%%%%%%%%%%%%%%%%%%%%%%%%%%%%%%%%%%%%%%%%%%%%%%%%%%%%%%%%%%%%%%%%%%%
\section{Introduction}
Symmetries play an important role in physics in general and in quantum field theory in particular.
The standard model (SM) as a low-energy effective theory has been very predictive and well tested, due to the symmetries satisfied by the theory - Lorentz invariance plus the $SU(3)_C\times SU(2)_L\times U(1)_Y$ gauge symmetry in addition to the discrete space-time symmetries like P and CP. However, it leaves many open questions for theoretical and cosmological issues that have not been solved yet. These include the following: inclusion of gravity in gauge theory, instability of the Higgs potential, cosmological puzzles of matter-antimatter asymmetry, dark matter, dark energy, and inflation, and flavor puzzle associated with the SM fermion mass hierarchies, their mixing patterns with the CP violating phases, and the strong CP problem.
  Moreover, there is no answer to the question: why there are three generations in the SM. The SM, therefore, cannot be the final answer.
So it is widely believed that the SM should be extended a more fundamental underlying theory. 
Neutrino mass and mixing is the first new physics beyond SM and adds impetus to solving the open questions in particle physics and cosmology. Moreover, a solution to the strong CP problem of QCD through Peccei-Quinn (PQ)\,\cite{Peccei-Quinn} mechanism\,\footnote{See, its related reports\,\cite{Cheng:1987gp}.} may hint a new extension of gauge theory realized in gauge/gravity duality\,\cite{Ahn:2016hbn}.
If nature is stringy, string theory, the only framework we have for a consistent theory with both quantum mechanics and gravity, should give insight into all such fundamental issues.
String theory when compactified to four dimensions can generically contain $G_F=$ {\it anomalous gauged $U(1)$} plus {\it non-Abelian finite symmetries}. In this regard, in order to construct a model with the open questions one needs more types of gauge symmetry beside the SM gauge theory. %with at least one more scalar field much heavier than the electroweak scale. 
One of simple approaches to a neat solution for those could be accommodated  by a type of symmetry based on seesaw\,\cite{Minkowski:1977sc} and Froggatt-Nielsen (FN)\,\cite{Froggatt:1978nt} frameworks, since it is widely believed that non-renormalizable operators in the effective theory should come from a more fundamental underlying renormalizable theory by integrating out the heavy degrees of freedom. Therefore, one can anticipate that there may exist some correlations between low energy and high energy physics; {\it e.g.} the flavored-axion\,\cite{Ahn:2016hbn} can easily fit into a string theoretic framework, and appear cosmologically as a form of cold dark matter. Even gravity (which is well-described by Einstein's general theory of relativity) lies outside the purview of the SM, once the gauged $U(1)$s are introduced in an extended theory, its mixed gravitational-anomaly should be free. And we assume that the heavy gauge bosons associated with the gauged $U(1)$s are decoupled, and thus in the model we consider the gauged $U(1)$s will be treated as the global $U(1)$s symmetries at low energy.  As shown in Ref.\,\cite{Ahn:2016hbn}, the FN mechanism formulated with global $U(1)$ flavor symmetry could be promoted from the string-inspired gauged $U(1)$ symmetry. Such flavored-PQ global symmetry $U(1)$ acts as a bridge for the flavor physics and string theory\,\cite{Ahn:2016hbn,Ahn:2016typ}. 
Flavor modeling on the non-Abelian finite group has been recently singled out as a good candidate to depict the flavor mixing patterns, {\it e.g}., Ref.\,\cite{Ahn:2016hbn, Feruglio:2007uu, Ahn:2014gva}, since it is preferred by vacuum configuration for flavor structure. Hence, flavored-PQ symmetry modeling extended to $G_F$ could be a powerful tool to resolve the open questions for particle physics and cosmology.

In this paper we present, by showing an extended flavored-PQ model which extend to a compact symmetry\,\footnote{Here the meaning of a `compact' symmetry is a symmetry that provides only requisite parameters it is not hard to disprove; for example, see the quark and lepton mass textures in Eqs.\,(\ref{Quark}) and (\ref{AxionLag2}) provided by the well-sewed supepotentials (\ref{lagrangian_q}) and (\ref{lagrangian2}) under the $SL_2(F_3)\times U(1)_X$ symmetry.} $G_F$ for new physics beyond SM, that the $U(1)$ mixed-gravitational anomaly cancellation is of central importance in constraining the fermion contents of a new chiral gauge theory, and the flavor structure of $G_F$ is\,\footnote{Here we assume that, below the scale associated with $U(1)_{X_i}$ gauge bosons, the gauged $U(1)_{X_i}$ leaves behind low-energy symmetries which are QCD anomalous global $U(1)_{X_i}$, see Eq.\,(\ref{gaugeU}).} strongly correlated with physical observables. So, finding the SM fermion mass spectra and their peculiar flavor mixing patterns in modeling is very important, since it is the first step toward establishing an effective low-energy Lagrangian of an extended theory. 
Unlike the $A_4$ symmetry containing one- and three-dimensional representations used in Refs.\cite{Ahn:2014gva, Ahn:2016hbn} the non-abelian discrete $SL_2(F_3)$ symmetry\,\cite{Feruglio:2007uu,slf3,aranda} contains two-dimensional representation in addition to one- and three-dimensional representations, in which the three dimensional representation is mainly responsible for the large leptonic mixing angles while the two dimensional representation is mainly to fit the quark masses and small mixing angles (especially the Cabbibo angle). Moreover, depending on the quantum number of flavored $U(1)_X$ the group $G_F$ can give different structures of quark and lepton mass texture. Together with $U(1)_X$ symmetry, such $SL_2(F_3)$ could make the model compact providing an economic mass texture (see Eq.\,(\ref{Quark})) for the quark mass spectra and mixings, especially, the Cabbibo angle. On the other hand, if one uses $A_4$ symmetry in the same framework, it is expected that there are uncontrollable redundant parameters in the quark mass textures which should be fine-tuned by hand to realize the quark mass spectra and mixings.
So taking $G_F=SL_2(F_3)\times U(1)_X$ may have a good advantage to compactly describe the peculiar mixing patterns of quarks and leptons including their masses. 
Contrary to Ref.\,\cite{Ahn:2018cau}, the present model provides another possibility of flavor modeling in virtue of the quantum number of $U(1)_X$, leading to completely different mass textures of quark and lepton. And in turn its results give an upper bound on QCD axion mass with different values of $\tan\beta$ in Eq.\,(\ref{tanpara}) and $g_{Aee}$ in Eq.\,(\ref{axion-electron0}), since axion to leptons and quarks couplings depend on structure of the quark and lepton sector. In this sense, if the astronomical constraint of star cooling\,\cite{Giannotti:2017hny} favored by the model in Ref.\,\cite{Ahn:2018cau}  is really responsible for the QCD axion, the present model will be ruled out. And it is expected that the upcoming NA62 experiment expected to reach the sensitivity of ${\rm Br}(K^+\rightarrow\pi^++A_i)<1.0\times10^{-12}$\,\cite{Fantechi:2014hqa} will soon rule out or favor the scenario in Ref.\,\cite{Ahn:2018cau}, while for the present model just gives an upper bound on the scale of PQ symmetry breakdown.

The rest of this paper is organized as follows. In Sec.\,II we set up a minimalistic SUSY model for quarks, leptons, and flavored-axions (and its combination QCD axion), which contains a $G_F=SL_2(F_3)\times U(1)_X$ symmetry for a compact description of new physics beyond SM. 
In Sec.\,III the $SL_2(F_3)\times U(1)_X$ symmetry-invariant superpotential for vacuum configurations is constructed and its vacuum structure is analyzed.
In Sec.\,IV we describe the Yukawa superpotential for quarks and flavored-axions and show that the SM quark masses and mixings could well be described by new expansion parameters defined under the $U(1)_X\times[gravity]^2$ anomaly-free condition. In turn, in order to show that the quark sector works well we perform a numerical simulation.
And we show that the constraint coming from the particle physics on rare decay $K^+\rightarrow \pi^++A_i$\,\cite{Wilczek:1982rv, Feng:1997tn, Ahn:2018cau} on the $U(1)_X$ symmetry breaking scale is much stronger than that from the astroparticle physics on QCD axion cooling of stars.
Along the line of quark sector, in Sec.\,V we show that the Yukawa superpotential for leptons and flavored-axions could well be flavor-structured, which gives testable predictions on the neutrino mass ordering, $\delta_{CP}$ and $\theta_{23}$. And we show that the $U(1)_X$ symmetry breaking scale can also be constrained via the astrophysical constraint on flavored-axion cooling of stars, but its constraint is smaller than that from $K^+\rightarrow\pi^++A_i$. % In order to fix the scale of PQ phase transition we take a testable scale, which gives model predictions on the axion mass and axion-photon coupling. 
What we have done is summarized in Sec.\,VI, and we provide our conclusions. In appendix we consider possible next-to-leading order corrections. 

%%%%%%%%%%%%%%%%%%%%%%%%%%%%%%%%%%%%%%%%%%%%%%%%%%%%%%%%%%%%%%%%%%%%%%%%%%%%
\section{The model setup}
Assume we have a SM gauge theory based on the $G_{\rm SM}=SU(3)_C\times SU(2)_L\times U(1)_Y$ gauge group, and that the theory has in addition a $G_F=SL_2(F_3)\times U(1)_X$ for a {\it compact} description of new physics beyond SM. Here the symmetry group of the double tetrahedron $SL_2(F_3)$\,\cite{slf3, aranda, Feruglio:2007uu}\,\footnote{The details of the $SL_2(F_{3})$ group are shown in Appendix\,\ref{SL2F3}.} is mainly for the peculiar flavor mixing patterns. Here we assume that the non-Abelian finite group $SL_2(F_3)$ could be realized in field theories on orbifolds and it is a subgroup of a gauge symmetry that can be protected from quantum-gravitational effects.
Since chiral fermions are certainly a main ingredient of the SM, the gauge- and gravitational-anomalies of the gauged $U(1)_X$  are\,\footnote{As shown in Refs.\,\cite{Ahn:2016typ,Ahn:2016hbn} with the well-defined Kahler potential based on type-IIB string theory, the author demonstrated that, while the two massive gauge bosons associated with the gauged $U(1)_{X_i}$ eat two degree of freedom, the other two axionic directions survive to low energies as the flavored-PQ axions, leaving behind low energy symmetries which are the QCD anomalous global $U(1)_{X_i}$.} generically present, making the theory inconsistent, where
\begin{eqnarray}
 U(1)_X\equiv U(1)_{X_1}\times U(1)_{X_2}\,.
  \label{gaugeU}
\end{eqnarray}
 Some requirements and constraints needed for the extended theory are: 
\begin{description}
\item[(i)] The mixed $G_{\rm SM}\times U(1)_{X_i}\times U(1)_{X_j}$ and cubic $U(1)_{X_i}\times[U(1)_{X_j}]^2$ anomalies should be cancelled by the Green-Schwarz (GS) mechanism\,\cite{Green:1984sg}. Hereafter the gauged $U(1)$ will be treated as the global $U(1)$ symmetry.  Note that the global symmetry $U(1)_X$ we consider is the remnant of the $U(1)_X$ gauge symmetry broken by the GS mechanism. Hence, the spontaneous breaking of $U(1)_X$ realizes the existence of the Nambu-Goldstone (NG) modes (called axions) and provides an elegant solution to the strong CP problem.
\item[(ii)] The non-vanishing anomaly coefficient of the quark sector $\{U(1)_{X_i}\times[gravity]^2\}_{\rm quark}$ constrains the quantity $\sum^{N_f}_{j}X_{\psi_j}$ in the gravitational instanton backgrounds (with $N_f$ generations well defined in the non-Abelian discrete group), and in turn whose quantity is congruent to the $U(1)_{X_i}\times[SU(3)_C]^2$ anomaly coefficient 
\begin{eqnarray}
 \delta^{\rm G}_k\delta^{ab}=2\sum_{\psi_i}X_{k\psi_i}{\rm Tr}(t^at^b)\,,
  \label{color_co}
\end{eqnarray}
 in the QCD instanton backgrounds, where the $t^a$ are the generators of the representation of $SU(3)$ to which Dirac fermion $\psi_i$ belongs with $X$-charge. Thanks to the two QCD anomalous $U(1)$ we have a relation\,\cite{Ahn:2014gva} 
\begin{eqnarray}
 |\delta^{\rm G}_1/\delta^{\rm G}_2|=|f_{a_1}/f_{a_2}|\,,
  \label{scale1}
\end{eqnarray}
indicating that the ratio of QCD anomaly coefficients is fixed by that of the decay constants $f_{a_i}$ of the flavored-axions $A_i$. Here $f_{a_i}$ set the flavor symmetry breaking scales, and their ratios appear in expansion parameters of the quark and lepton mass spectra (see Eqs.\,(\ref{expan_0}) and (\ref{expan_1})).  As studied in Refs.\,\cite{Ahn:2014gva, Ahn:2016hbn}, in the so-called flavored-PQ models the scale of PQ symmetry breakdown is congruent to the seesaw scale via Eq.\,(\ref{scale1}), which could well be fixed\,\footnote{If one takes seriously the hints from axion cooling of stars in Refs.\,\cite{Isern:2008nt, Bertolami:2014wua}, one can fix the scale of PQ symmetry breakdown congruent to the seesaw scale\,\cite{Ahn:2016hbn}.} and/or constrained through the constraints and/or hints coming from astroparticle physics on axion cooling of stars with the fine-structure of axion to electron $\alpha_{Aee}<6\times10^{-27}$\,\cite{Bertolami:2014wua}, $4.1\times10^{-28}\lesssim\alpha_{Aee}\lesssim3.7\times10^{-27}$\,\cite{Bertolami:2014wua}, and the coupling of axion to neutron $g_{Ann}<8\times10^{-10}$\,\cite{Sedrakian:2015krq} etc. as well as the constraints coming from particle physics on rare flavor violating decay processes induced by the flavored-axions ${\rm Br}(K^+\rightarrow\pi^+A_i)<7.3\times10^{-11}$\,\cite{Adler:2008zza} and ${\rm Br}(\mu\rightarrow e\,\gamma\,A_i)\lesssim1.1\times10^{-9}$\,\cite{Bolton:1988af} etc..
\item[(iii)]The mixed-gravitational anomaly $U(1)_X\times[gravity]^2$ must be cancelled to consistently couple gravity to matter charged under $U(1)_X$. Since a heavy Majorana neutrino (necessary to implement the seesaw and PQ mechanisms, simultaneously) with $U(1)_{X_1}$ charge $X_1/2$ does not have a vanishing $U(1)_{X_1}\times[gravity]^2$ anomaly, its anomaly should be cancelled by another contribution of $U(1)_{X_2}\times[gravity]^2$ anomaly. Hence, the $U(1)_X$ charges of SM fermions and new fermions including heavy Majorana neutrinos 
must be commensurate through the $U(1)_X\times[gravity]^2$ anomaly satisfying a condition
\begin{eqnarray}
 k_1\,X_1/2=k_2\,X_2
  \label{cond}
\end{eqnarray}
where\,\footnote{For $-k_1=k_2=1$ in Ref.\,\cite{Ahn:2016hbn}, additional Majorana fermions are introduced to satisfy the $U(1)_{X}\times[gravity]^2$ anomaly free-condition. Note that, however, in general, $k_2/k_1\neq$ {\it integer}.} $k_i$ $(i=1,2)$ are nonzero integers, which is a conjectured relationship between two anomalous $U(1)$s. The $U(1)_{X_i}$ is broken down to its discrete subgroup $Z_{N_i}$ in the backgrounds of QCD instanton, and the quantities $N_i$ ({\it nonzero integers}) associated to the axionic domain-wall are given by
\begin{eqnarray}
 \Big|\frac{\delta^{\rm G}_1}{X_1/2k_2}\Big|=N_1\,,\quad \Big|\frac{\delta^{\rm G}_2}{X_2/k_1}\Big|=N_2\,.
  \label{dw}
\end{eqnarray}
Then, from Eqs.\,(\ref{cond}) and (\ref{dw}) one obtains $|\delta^{\rm G}_1|=N_1$ and $|\delta^{\rm G}_2|=N_2$.
Clearly, in the QCD instanton backgrounds if $N_1$ and $N_2$ are relative prime, there is no $Z_{N_{\rm DW}}$ discrete symmetry and therefore no domain wall problem\,\footnote{Note that, in the present model, since the non-Abelian finite symmetry $SL_2(F_3)$ is broken completely by higher order effects, there is no residual symmetry; so, there is no room for a spontaneously  broken discrete symmetry to lead to domain-wall problem.}. 
Now, we will see that the domain-wall condition with the $U(1)_{X}\times[gravity]^2$ anomaly free-condition is dependent on the $U(1)_X$ charged quark and lepton flavors. 
Eq.\,(\ref{color_co}) can be expressed $\delta^{\rm G}_1=\alpha\,X_1$ and $\delta^{\rm G}_2=\omega\,X_2$, where $\alpha$ and $\omega$ are some integer numbers. To make sure that no axionic domain-wall problem occurs, the following two conditions are required: (i) {\it The numbers $\alpha$ and $\omega$ coming from $U(1)_X$ charged quark flavors should be `relative prime'}. If the quantum numbers $X_1$ and $X_2$ are given by $-2p$ and $-q$, respectively,  from Eq.\,(\ref{cond}) one obtains $k_1\,p=k_2\,q$. So the number $k_i$ coming from the $U(1)_{X}\times[gravity]^2$ anomaly-free condition depends on both the $U(1)_X$ charged quark and lepton flavors. Then, Eq.\,(\ref{dw}) is expressed as 
\begin{eqnarray}
 N_1=|\delta^{\rm G}_1|=2|\alpha|\,k_2\,,\quad N_2=|\delta^{\rm G}_2|=|\omega|\,k_1\,.
  \label{dwahn}
\end{eqnarray}
(ii) Hence, {\it the number $k_2$ should be relative prime with $|\omega|$ and $k_1$, as well as the number $k_1$ should not be a multiple of $2$ and should be relative prime with $|\alpha|$}. 

Consequently\,\footnote{Of course, one can consider the cases of the domain-wall number $N_{\rm DW}>1$ if the PQ phase transition occurred during (or before) inflation.}, under the $U(1)_{X}\times[gravity]^2$ anomaly-free condition, to make sure that no axionic domain-wall problem occurs in a theory one could introduce additional $U(1)_X$ charged Majorana fermions and/or could assign well flavor-structured $U(1)_X$ quantum numbers to fermion contents that can protect $k_1$ to be a multiple of 2.
\end{description}

As we shall see later, even though the integer $k_i$ depends on both the $U(1)_X$ charged quark and lepton flavors, it does not play the role of constraining the QCD axion decay constant $F_A=f_{a_i}/\delta^{\rm G}_{i}\sqrt{2}$ through physical processes induced by flavored-axions in the flavored-PQ models. On the other hand, those physical processes are constrained by $2\alpha$ and $\omega$  coming from the QCD instanton background.

 Along this line, the $G_F$ quantum number of the field contents is assigned in the following two ways: (a) in a way that the $SL_2(F_3)$ that compactly depict the Cabbibo-Kobayashi-Maskawa (CKM) for quark mixings and  Pontecorvo-Maki-Nakagawa-Sakata (PMNS) for leptonic mixings requires a desired vacuum configuration, and
(b) the $U(1)_X$ mixed-gravitational anomaly-free condition with the SM flavor structure demands additional  Majorana fermions as well as no axionic domain-wall problem.

%%%%%%%%%%%%%%%%%%%%%%%%%%%%%%%%%%%%%%%%%%%%%%%%%%%%%%%%%%%%%%%%%%%%%%%%%%%%
\section{Vacuum configuration}
In this section, the $SL_2(F_3)\times U(1)_X$ symmetry-invariant superpotential for vacuum configurations is constructed and its vacuum structure is analyzed.
First we present the representations of the field contents responsible for vacuum configuration.
Apart from the usual two Higgs doublets $H_{u,d}$ responsible for electroweak symmetry breaking, which are invariant under $SL_2(F_3)$ ({\it i.e.} flavor singlets $\mathbf{1}$), the scalar sector is extended via two types of new scalar multiplets, flavon fields responsible for the spontaneous breaking of the flavor symmetry $\Phi_{T},\Phi_{S},\Theta,\tilde{\Theta}, \eta, \Psi, \tilde{\Psi}$ that are $G_{\rm SM}$-singlets and driving fields $\Phi^{T}_{0},\Phi^S_{0},\eta_0, \Theta_{0},\Psi_{0}$ that are to break the flavor group along required vacuum expectation value (VEV) directions and to allow the flavons to get VEVs, which couple only to the flavons: we take the flavon fields $\Phi_{T},\Phi_{S}$ to be $SL_2(F_3)$ triplets, $\eta$ to be a $SL_2(F_3)$ doublet ($\mathbf{2}'$ representation), and $\Theta,\tilde{\Theta},\Psi,\tilde{\Psi}$ to be $SL_2(F_3)$ singlets ($\mathbf{1}$ representation), respectively, that are $G_{\rm SM}$-singlets, and driving fields $\Phi_{0}^{T},\Phi_{0}^{S}$ to be $SL_2(F_3)$ triplets, $\eta_0$ to be a $SL_2(F_3)$ doublet ($\mathbf{2}''$ representation) and $\Theta_{0}, \Psi_{0}$ to be $SL_2(F_3)$ singlets.
The flavored-PQ symmetry $U(1)_X$ is composed of two anomalous symmetries $U(1)_{X_1}\times U(1)_{X_2}$ generated by the charges $X_1\equiv-2p$ and $X_2\equiv-q$. The flavon fields $\{\Phi_S,\Theta,\tilde{\Theta}\}$ are $X_1$ charged, and $\{\Phi^S_0, \Theta_0\}$ are $ -2X_1$ charged, respectively, under $U(1)_{X_1}$; the field $\Psi$ ($\tilde{\Psi})$ is $X_2 (-X_2)$ charged under $U(1)_{X_2}$. For vacuum stability and a desired vacuum alignment solution, we enforce $\{\Phi_T, \eta\}$ to be neutral under $U(1)_X$. And the others $H_{u,d}$, $\Phi^T_0, \eta_0$, and $\Psi_0$ are neutral under $U(1)_X$. % not to have an axionic domain-wall problem. And the others are neutral under $U(1)_X$.
 Moreover, the superpotential $W$ in the theory is uniquely determined by the $U(1)_R$ symmetry, containing the usual $R$-parity as a subgroup: $\{matter\,fields\rightarrow e^{i\xi/2}\,matter\,fields\}$ and $\{driving\,fields\rightarrow e^{i\xi}\,driving\,fields\}$, with $W\rightarrow e^{i\xi}W$, whereas flavon and Higgs fields remain invariant under an $U(1)_R$ symmetry. As a consequence of the $R$ symmetry, the other superpotential term $\kappa_{\alpha}L_{\alpha}H_{u}$ and the terms violating the lepton and baryon number symmetries are not allowed. In addition, dimension 6 supersymmetric operators like $Q_{i}Q_{j}Q_{k}L_{l}$ ($i,j,k$ must not all be the same) are not allowed either, and stabilizing proton. %It is likely that an exact continuous global symmetry is violated by quantum gravitational effects\,\cite{Krauss:1988zc}. 
 Here the global $U(1)$ symmetry is a remnant of the broken $U(1)$ gauge symmetry which can connect string theory with flavor physics\,\cite{Ahn:2016typ, Ahn:2016hbn} (see also\,\cite{somorelated}).

Under $SL_2(F_{3})\times U(1)_{X}\times U(1)_{R}$, representations of the driving, flavon, and Higgs fields are summarized as in Table\,\ref{DrivingRef}.
%\begin{center}
\begin{table}[h]
\caption{\label{DrivingRef} Representations of the driving, flavon, and Higgs fields under $SL_2(F_{3})\times U(1)_{X}\times U(1)_R$. Here $U(1)_X\equiv U(1)_{X_1}\times U(1)_{X_2}$ symmetries which are generated by the charges $X_1=-2p$ and $X_2=-q$.}
\begin{ruledtabular}
\begin{tabular}{ccccccccccccccccc}
Field &$\Phi^{T}_{0}$&$\Phi^{S}_{0}$&$\Theta_{0}$&$\Psi_{0}$&$\eta_{0}$&\vline\vline&$\Phi_{S}$&$\Phi_{T}$&$\Theta$&$\tilde{\Theta}$&$\Psi$&$\tilde{\Psi}$&$\eta$&\vline\vline&$H_{d}$&$H_{u}$\\
\hline
$SL_2(F_{3})$&$\mathbf{3}$&$\mathbf{3}$&$\mathbf{1}$&$\mathbf{1}$&$\mathbf{2}''$&\vline\vline&$\mathbf{3}$&$\mathbf{3}$&$\mathbf{1}$&$\mathbf{1}$&$\mathbf{1}$&$\mathbf{1}$&$\mathbf{2}'$&\vline\vline&$\mathbf{1}$&$\mathbf{1}$\\
$U(1)_{X}$&$0$&$4p$&$4p$&$0$&$0$&\vline\vline&$-2p$&$0$&$-2p$&$-2p$&$-q$&$q$&$0$&\vline\vline&$0$&$0$\\
$U(1)_R$&$2$&$2$&$2$&$2$&$2$&\vline\vline&$0$&$0$&$0$&$0$&$0$&$0$&$0$&\vline\vline&$0$&$0$\\
%$SU(2)\times U(1)_Y$&$1_{0}$&$1_{0}$&$1_{0}$&$1_{0}$&$1_{0}$&$1_{0}$&$1_{0}$\\
\end{tabular}
\end{ruledtabular}
\end{table}
%\end{center}
The superpotential depending on the driving fields, invariant under $G_{\rm SM}\times U(1)_R\times G_F$, reads at leading order
\begin{eqnarray}
 W_v&=&\Phi^T_0(\mu_T\Phi_T+g_T\Phi_T\Phi_T)+\Phi^S_0(g_1\Phi_S\Phi_S+g_2\tilde{\Theta}\Phi_S)+\eta_0(\mu_\eta\eta+g_\eta\eta\Phi_T)\nonumber\\
 &+&\Theta_0(g_3\Phi_S\Phi_S+g_4\Theta\Theta+g_5\Theta\tilde{\Theta}+g_6\tilde{\Theta}\tilde{\Theta})+g_7\Psi_0(\Psi\tilde{\Psi}-\mu^2_\Psi)
 +g_8\Phi^T_0\eta\eta\,,
  \label{super_d}
\end{eqnarray}
where higher dimensional operators are neglected, and $\mu_{i=T,\Psi, \eta}$ are dimensional parameters and $g_{T, \eta}, g_{1,...,8}$ are dimensionless coupling constants. Note here that the model implicitly has two $U(1)_{X}\equiv U(1)_{X_1}\times U(1)_{X_2}$ symmetries which are generated by the charges $X_{1}=-2p$ and $X_{2}=-q$.
The fields $\Psi$ and $\tilde{\Psi}$ charged by $-q,q$, respectively, are ensured by the $U(1)_{X}$ symmetry extended to a complex $U(1)$ due to the holomorphy of the supepotential. So, the PQ scale $\mu_\Psi=\sqrt{v_{\Psi}v_{\tilde{\Psi}}/2}$ corresponds to the scale of spontaneous symmetry breaking of the $U(1)_{X_2}$
symmetry.  
Since there is no fundamental distinction between the singlets $\Theta$ and $\tilde{\Theta}$ as indicated in Table\,\ref{DrivingRef}, we are free to define $\tilde{\Theta}$ as the combination that couples to $\Phi^{S}_{0}\Phi_{S}$ in the superpotential $W_{v}$\,\cite{Altarelli:2005yp}. 
At the leading order the usual superpotential term $\mu H_{u}H_{d}$ is not allowed, while at the leading order the operator driven by $\Psi_0$ and at the next leading order the operators driven by $\Phi^T_0$ and $\eta_0$ are allowed
 \begin{eqnarray}
  g_{\Psi_0}\Psi_0\,H_uH_d+\frac{g_{T_0}}{\Lambda}(\Phi^{T}_{0}\Phi_{T})_{{\bf 1}}H_{u}H_{d}+\frac{g_{\eta_0}}{\Lambda}(\eta_{0}\,\eta)_{{\bf 1}}H_{u}H_{d}\,,
 \label{muterm}
 \end{eqnarray}
which is to promote the effective $\mu$-term $\mu_{\rm eff}\equiv g_{\Psi_0}\langle\Psi_{0}\rangle+g_{T_0}\langle\Phi^{T}_{0}\rangle\, v_{T}/(\sqrt{2}\Lambda)+g_{\eta_0}\langle\eta_{0}\rangle\, v_{\eta}/(\sqrt{2}\Lambda)$ of the order of $m_S$, $m_{S}\,v_{T}/\Lambda$, and  $m_{S}\,v_{\eta}/\Lambda$ (here $\langle\Psi_{0}\rangle$, $\langle\Phi^{T}_{0}\rangle$, and $\langle\eta_{0}\rangle$: the VEVs of the scalar components of the driving fields, $m_{S}$: soft SUSY breaking mass). It is interesting that at the leading order the electroweak scale does not mix with the potentially large scales, the VEVs of the scalar components of the flavon fields, $v_{S},v_{T},v_{\Theta}$, $v_{\eta}$ and $v_{\Psi}$. Actually, in the model once the scale of breakdown of $U(1)_X$ symmetry is fixed by the constraints coming from astrophysics and particle physics, the other scales are automatically fixed by the flavored model structure. 
And it is clear that at the leading order the scalar supersymmetric $W(\Phi_{T}\Phi_{S})$ terms are absent due to different $U(1)_{X}$ quantum numbers, which is crucial for relevant vacuum configuration in the model to produce compactly the present lepton and quark mixing angles.
Now we consider how a desired vacuum configuration for compact description of quark and lepton mixings could be derived.  
In SUSY limit, the vacuum configuration is obtained by the $F$-terms of all fields being required to vanish. The vacuum alignments of the flavons $\Phi_{T}$ and $\eta$ are determined by
 \begin{eqnarray}
 \frac{\partial W_{v}}{\partial\Phi^{T}_{01}}&=&\mu_T\,\Phi_{T1}+\frac{2g_T}{3}\left(\Phi^{2}_{T1}-\Phi_{T2}\Phi_{T3}\right)+ig_8\,\eta^2_1=0\,,\nonumber\\
 \frac{\partial W_{v}}{\partial\Phi^{T}_{02}}&=&\mu_T\,\Phi_{T3}+\frac{2g_T}{3}\left(\Phi^{2}_{T2}-\Phi_{T1}\Phi_{T3}\right)+g_8\,(1-i)\eta_1\eta_2=0\,,\nonumber\\
 \frac{\partial W_{v}}{\partial\Phi^{T}_{03}}&=&\mu_T\,\Phi_{T2}+\frac{2g_T}{3}\left(\Phi^{2}_{T3}-\Phi_{T1}\Phi_{T2}\right)+g_8\,\eta^2_2=0\, \label{potential1}
 \end{eqnarray}
  \begin{eqnarray}
 \frac{\partial W_{v}}{\partial\eta_{01}}&=&\mu_\eta\,\eta_{2}+\frac{5g_\eta}{6}\Big(\frac{1-i}{2}\eta_{2}\Phi_{T1}+i\eta_1\Phi_{T3}\Big)=0\,\nonumber\\
 \frac{\partial W_{v}}{\partial\eta_{02}}&=&-\mu_\eta\,\eta_{1}+\frac{5g_\eta}{6}\Big(\frac{1-i}{2}\eta_{1}\Phi_{T1}+i\eta_2\Phi_{T2}\Big)=0\,
 \label{potential01}
 \end{eqnarray}
From this set of five equations, we can obtain the supersymmetric vacua for $\Phi_{T}$ and $\eta$
 \begin{eqnarray}
\langle\Phi_{T}\rangle&=&\Big(\frac{v_{T}}{\sqrt{2}},\,0,\,0\Big)\,,\qquad \text{with}\,\, \mu_{T}=-g_T\frac{\sqrt{2}}{3}v_T-i\frac{g_8}{\sqrt{2}}\frac{v^2_\eta}{v_T}\,,\nonumber\\
\langle\eta\rangle&=&\Big(\pm\frac{v_{\eta}}{\sqrt{2}},\,0\Big)\,,\qquad~\, \text{with}\,\, \mu_{\eta}=g_\eta\,\frac{v_T}{\sqrt{2}}\frac{5(1-i)}{12}\,,
 \label{vevdirection1}
 \end{eqnarray}
where $g_T$ and $g_\eta$ are dimensionless couplings, and $v_T$ and $v_\eta$ are not determined.
The minimization equations for the vacuum configuration of $\Phi_{S}$ and $(\Theta,\tilde{\Theta})$ are given by
 \begin{eqnarray}
 \frac{\partial W_{v}}{\partial\Phi^{S}_{01}}&=&\frac{2g_{1}}{3}\left(\Phi_{S1}\Phi_{S1}-\Phi_{S2}\Phi_{S3}\right)+g_{2}\Phi_{S1}\tilde{\Theta}=0\,,\nonumber\\
 \frac{\partial W_{v}}{\partial\Phi^{S}_{02}}&=&\frac{2g_{1}}{3}\left(\Phi_{S2}\Phi_{S2}-\Phi_{S1}\Phi_{S3}\right)+g_{2}\Phi_{S3}\tilde{\Theta}=0\,,\nonumber\\
 \frac{\partial W_{v}}{\partial\Phi^{S}_{03}}&=&\frac{2g_{1}}{3}\left(\Phi_{S3}\Phi_{S3}-\Phi_{1}\Phi_{S2}\right)+g_{2}\Phi_{S2}\tilde{\Theta}=0\,,\nonumber\\
 \frac{\partial W_{v}}{\partial\Theta_{0}}&=&g_{3}\left(\Phi_{S1}\Phi_{S1}+2\Phi_{S2}\Phi_{S3}\right)+g_{4}\Theta^{2}+g_{5}\Theta\tilde{\Theta}+g_{6}\tilde{\Theta}^{2}=0\,.
 \label{potential2}
 \end{eqnarray}
And from  Eq.\,(\ref{potential2}), we can get the supersymmetric vacua for the fields $\Phi_{S},\Theta,\tilde{\Theta}$
 \begin{eqnarray}
 \langle\Phi_{S}\rangle=\frac{1}{\sqrt{2}}\left(v_{S},v_{S},v_{S}\right)\,,\quad\langle\Theta\rangle=\frac{v_{\Theta}}{\sqrt{2}}\,,\quad\langle\tilde{\Theta}\rangle=0\,,\qquad\text{with}\,\,v_{\Theta}=v_{S}\sqrt{-3\frac{g_{3}}{g_{4}}}\,,
 \label{vevdirection2}
 \end{eqnarray}
where $v_{\Theta}$ is undetermined. As can be seen in Eq.\,(\ref{vevdirection2}), the VEVs $v_{\Theta}$ and $v_{S}$ are naturally of the same order of magnitude (here the dimensionless parameters $g_{3}$ and $g_{4}$ are the same order of magnitude).
Finally, the minimization equation for the vacuum configuration of $\Psi$ is given by
 \begin{eqnarray}
 \frac{\partial W_{v}}{\partial\Psi_{0}}&=&g_{7}(\Psi\tilde{\Psi}-\mu^{2}_{\Psi})=0\,,
 \label{potential3}
 \end{eqnarray}
where $\mu_{\Psi}$ is the $U(1)_{X}$ breaking scale and $g_{7}$ is a dimensionless coupling.
From the above equation we can get the supersymmetric vacua for the fields $\Psi,\tilde{\Psi}$
 \begin{eqnarray}
 \langle\Psi\rangle=\langle\tilde{\Psi}\rangle=\frac{v_{\Psi}}{\sqrt{2}}\,.
 \label{vevdirection3}
 \end{eqnarray}
Note that, once the scale of breakdown of $U(1)_X$ symmetry is fixed, all the other scales of VEVs are determined by the present flavor structured model.
%From the minimization conditions of the $F$-term scalar potential with respect to\,\footnote{The vacuum configuration of the driving fields is not relevant in this work. And  we will not consider seriously the corrections to the VEVs due to higher dimensional operators contributing to Eq.\,(\ref{super_d}) since their effects are expected to be only few percents level, see Appendix\,\ref{corre}.} $\Phi_T$, $\Phi_S$, $\eta$, $\tilde{\Theta}$, $\Psi$, and $\tilde{\Psi}$,  we see that the global minima of the potential are located at Eqs.\,(\ref{vevdirection1}), (\ref{vevdirection2}) and (\ref{vevdirection3}).
%We take a phenomenologically viable vacuum configuration:
%{\small\begin{eqnarray}
% &&\langle\Phi_T\rangle=\frac{v_{T}}{\sqrt{2}}(1,0,0)\,,\qquad\qquad\langle\Phi_S\rangle=\frac{v_{S}}{\sqrt{2}}(1,1,1)\,,\qquad\qquad\langle\eta\rangle=\frac{v_{\eta}}{\sqrt{2}}(1,0)\,,\nonumber\\
% &&\qquad\quad\langle\Psi\rangle=\langle\tilde{\Psi}\rangle=\frac{v_\Psi}{\sqrt{2}}\,, \qquad\qquad\langle\Theta\rangle=\frac{v_\Theta}{\sqrt{2}}\,,\qquad\qquad \langle\tilde{\Theta}\rangle=0\,,
%  \label{vevchi}
%\end{eqnarray}}
%where $v_\Psi=v_{\tilde{\Psi}}$ and $\kappa=v_S/v_\Theta$  in SUSY limit. 
As can be seen in Eqs.\,(\ref{vevdirection2}) and (\ref{vevdirection3}), in the SUSY limit there exist flat directions along which the scalar fields $\Phi_{S}, \Theta$ and $\Psi,\tilde{\Psi}$ do not feel the potential.
The SUSY-breaking effect lifts up the flat directions and corrects the VEV of the driving fields, leading to soft SUSY-breaking mass terms (here we do not specify a SUSY breaking mechanism in this work).

The flavon field ${\cal F}$ charged under $U(1)_X$ is a scalar field which acquires a VEV and breaks spontaneously the flavored-PQ symmetry $U(1)_{X}$. In order to extract NG modes resulting from spontaneous breaking of $U(1)_{X}$ symmetry, we set the decomposition of complex scalar fields as follows\,\footnote{Note that the massless modes are not contained in the $\tilde{\Theta},\Phi^{S}_{0},\Theta_{0}$ fields in supersymmetric limit.}
 \begin{eqnarray}
  &&\Phi_{Si}=\frac{e^{i\frac{\phi_{S}}{v_{S}}}}{\sqrt{2}}\left(v_{S}+h_{S}\right)\,,\qquad\qquad\quad\,\Theta=\frac{e^{i\frac{\phi_{\theta}}{v_{\Theta}}}}{\sqrt{2}}\left(v_{\Theta}+h_{\Theta}\right)\,,\nonumber\\
&&\Psi=\frac{v_{\Psi}}{\sqrt{2}}e^{i\frac{\phi_{\Psi}}{v_{g}}}\left(1+\frac{h_{\Psi}}{v_{g}}\right)\,,\qquad\qquad\,\tilde{\Psi}=\frac{v_{\tilde{\Psi}}}{\sqrt{2}}e^{-i\frac{\phi_{\Psi}}{v_{g}}}\left(1+\frac{h_{\tilde{\Psi}}}{v_{g}}\right)\,,
  \label{NGboson}
 \end{eqnarray}
in which we have set $\Phi_{S1}=\Phi_{S2}=\Phi_{S3}\equiv\Phi_{Si}$ in the supersymmetric limit, and $v_{g}=\sqrt{v^2_{\Psi}+v^2_{\tilde{\Psi}}}$. And the NG modes $A_1$ and $A_2$ are expressed as\,\cite{Ahn:2016hbn}
 \begin{eqnarray}
  A_1=\frac{v_{S}\,\phi_{S}+v_{\Theta}\,\phi_{\theta}}{\sqrt{v^{2}_{S}+v^{2}_{\Theta}}}\,,\qquad A_{2}=\phi_{\Psi}
 \end{eqnarray}
with the angular fields $\phi_{S}$, $\phi_{\theta}$ and $\phi_{\Psi}$.

 %%%%%%%%%%%%%%%%%%%%%%%%%%%%%%%%%%%%%%%%%%%%%%%%%%%%%%%%%%%%%%%%%%%%%%%%%%%%
\section{Quarks and flavored-Axions}
Let us impose $SL_2(F_3)\times U(1)_X$ quantum numbers on SM quarks in a way that quark masses and mixings are well described as well as no axionic domain-wall problem occurs\,\footnote{See Appendix\,\ref{axi_do}.}.

Under $SL_2(F_3)\times U(1)_X$, we assign the left-handed quark $SU(2)_{L}$ doublets denoted as $Q_{1}$, $Q_{2}$ and $Q_{3}$ to the $({\bf 1}, 4p+4q)$, $({\bf 1}', 2p+2q)$ and $({\bf 1}'', 0)$, respectively, while the right-handed up-type quark $SU(2)_{L}$ singlets are assigned as ${\cal U}^{c}=\{u^{c}, c^{c}\}$ and $t^{c}$ to the $({\bf 2}', -q-2p)$ and $({\bf 1}', 0)$, respectively, and the right-handed down-type quarks ${\cal D}^{c}=\{d^{c}, s^{c}\}$ and $b^c$ to the $({\bf 2}', -3q-2p)$ and $({\bf 1}', -q)$, respectively.
Under $SL_2(F_3)\times U(1)_{X}$ with $U(1)_R=+1$, the quantum numbers of the SM quark fields are summarized as in Table\,\ref{reps_q}.
%\begin{center}
\begin{table}[h]
\caption{\label{reps_q} Representations of the quark fields under $SL_2(F_3)\times U(1)_{X}$ with $U(1)_R=+1$.}
\begin{ruledtabular}
\begin{tabular}{cccc}
Field &$Q_{1},~Q_{2},~Q_{3}$&${\cal D}^c, ~b^c$&${\cal U}^c, ~t^c$\\
\hline
$SL_2(F_3)$&$\mathbf{1}$, $\mathbf{1}'$, $\mathbf{1^{\prime\prime}}$&$\mathbf{2}'$, $\mathbf{1}'$&$\mathbf{2}'$, $\mathbf{1^{\prime}}$\\
$U(1)_{X}$&$4p+4q,~2p+2q,~0$ &$-3q-2p$, ~$-q$&$-q-2p,~0$\\
%$U(1)_R$&$1$ &~$1$~&~$1$\\
%$SU(2)\times U(1)_Y$&$2_{-1}$&$2_\frac{1}{3}$&$1_{\frac{4}{3}}$&$1_{-\frac{2}{3}}$&$1_{-2}$&$1_{0}$&$2_{1}$&$2_{-1}$\\
\end{tabular}
\end{ruledtabular}
\end{table}
%\end{center} 
The $U(1)_{X}$ invariance forbids renormalizable Yukawa couplings for the light families, but would allow them through effective nonrenormalizable couplings suppressed by $({\cal F}/\Lambda)^n$ with some positive integer $n$. Here $\Lambda$, above which there exists unknown physics, is the scale of flavor dynamics, and is associated with heavy states which are integrated out.
The Yukawa superpotential for quark sector invariant under $G_{\rm SM}\times G_F\times U(1)_R$ is given by
%\begin{widetext}
{\begin{eqnarray}
 W_q &=&
  \hat{y}_{t}\,t^cQ_{3}H_u
  +y_{c}\,(\eta {\cal U}^c)_{{\bf 1}''}Q_2\frac{H_u}{\Lambda}+y_{u}\,[(\eta {\cal U}^c)_{\bf 3}\Phi_S]_{\bf 1}Q_1\frac{H_u}{\Lambda^2}\nonumber\\
  &+& y_{b}\,b^cQ_{3}H_d
  +y_{s}\,(\eta {\cal D}^c)_{{\bf 1}''}Q_2\frac{H_d}{\Lambda}+Y_{s}\,b^cQ_2(\Phi_S\Phi_T)_{{\bf 1}'}\frac{H_d}{\Lambda^2}+y_{d}\,[(\eta {\cal D}^c)_{\bf 3}\Phi_S]_{{\bf 1}}Q_{1}\frac{H_d}{\Lambda^2}\nonumber\\
  &+& Y_{d}\,b^cQ_1(\Phi_S\Phi_S)_{{\bf 1}''}\frac{H_d}{\Lambda^2}
  + \tilde{y}_{d}\,[(\eta {\cal D}^c)_{\bf 3}\Phi_T]_{{\bf 1}}Q_{1}\frac{H_d}{\Lambda^2}\,,
 \label{lagrangian_q}
 \end{eqnarray}}
%\end{widetext}
where the hat Yukawa coupling denotes order of unity {\it i.e.}, $1/\sqrt{10}\lesssim|\hat{y}|\lesssim\sqrt{10}$, and
\begin{eqnarray}
y_c &=&\hat{y}_c\Big(\frac{\Psi}{\Lambda}\Big)\,,\qquad y_u=\hat{y}_u\Big(\frac{\Psi}{\Lambda}\Big)^3\,,\qquad y_b=\hat{y}_b
\Big(\frac{\tilde{\Psi}}{\Lambda}\Big)\,,~~\qquad y_s=\hat{y}_s\Big(\frac{\tilde{\Psi}}{\Lambda}\Big)\nonumber\\
Y_s&=&\hat{Y}_s\Big(\frac{\Psi}{\Lambda}\Big)\,,\qquad y_d=\hat{y}_d\Big(\frac{\Psi}{\Lambda}\Big)\,, ~\qquad Y_d=\hat{Y}_d\Big(\frac{\Psi}{\Lambda}\Big)^{3}\,,\qquad\tilde{y}_d=\hat{\tilde{y}}_d\Big(\frac{\Psi}{\Lambda}\Big)\Big(\frac{\Theta}{\Lambda}\Big)\,.
\end{eqnarray}
Higher dimensional operators driven by $\Phi_T$ and $\eta$ fields, {\it e.g.} $\tilde{y}_{c}[(\eta {\cal U}^c)_{\bf 3}\Phi_T]_{{\bf 1}''}Q_2\frac{H_u}{\Lambda^2}$  with $\tilde{y}_{c}=\hat{\tilde{y}}_{c}(\Psi/\Lambda)$ is neglected here, but will be included in numerical calculation.

Once the scalar fields $\Phi_{S}, \Theta, \tilde{\Theta},\Psi$ and $\tilde{\Psi}$ get VEVs, the flavored $U(1)_{X}$ symmetry is spontaneously broken\,\footnote{If the symmetry $U(1)_{X}$ is broken spontaneously, the massless modes $A_1$ of the scalar $\Phi_{S}$ (and/or $\Theta$) and $A_{2}$ of the scalar $\Psi(\tilde{\Psi})$ appear as phases.}. 
And at energies below the electroweak scale, all quarks and leptons obtain masses. 
The relevant quark interaction terms with chiral fermions is given by 
 \begin{eqnarray}
  -{\cal L}^{q}_{WY} &=&
  \overline{q^{u}_{R}}\,\mathcal{M}_{u}\,q^{u}_{L}+\overline{q^{d}_{R}}\,\mathcal{M}_{d}\,q^{d}_{L} +\frac{g}{\sqrt{2}}W^+_\mu\overline{q^u_{L}}\gamma^\mu\,q^d_{L}+\text{h.c.}\,,
  \label{AxionLag1}
 \end{eqnarray}
where $q^{u}=(u,c,t)$, $q^{d}=(d,s,b)$, and $g$ is the SU(2) coupling constant.
With the desired direction of Eqs.\,(\ref{vevdirection1}, \ref{vevdirection2}, \ref{vevdirection3})\,\footnote{Here we took $\langle\eta\rangle=\frac{v_\eta}{\sqrt{2}}(+1,0)$.} the up(down)-type quark mass matrices in the above Lagrangian (\ref{AxionLag1}) read\,\footnote{Even there seem to have vacuum corrections to the leading order picture in Eq.\,(\ref{Quark}), {\it e.g.} $-\hat{y}_s\frac{\delta v_{\eta_2}}{\Lambda}\nabla_\Psi\,d^cQ_2\,H_d$ and $-\hat{y}_c\frac{\delta v_{\eta_2}}{\Lambda}\nabla_\Psi\,u^cQ_2\,H_u$, by the higher-dimensional operators in the driving superpotential Eq.\,(\ref{Npotential}), one can make their contributions vanishing or small enough.}
 \begin{eqnarray}
 &\mathcal{M}_{u}={\left(\begin{array}{ccc}
 iy_{u}\nabla_\eta\nabla_S\,e^{i(\frac{A_1}{v_{\cal F}}+3\frac{A_2}{v_g})} & 0 & 0 \\
 \frac{1-i}{2}y_u\nabla_\eta\nabla_S\,e^{i(\frac{A_1}{v_{\cal F}}+3\frac{A_2}{v_g})}  & y_{c}\nabla_\eta\,e^{i\frac{A_2}{v_g}} & 0 \\
 0 & 0 & \hat{y}_t
 \end{array}\right)}v_u\,,\nonumber\\
 &\mathcal{M}_{d}={\left(\begin{array}{ccc}
(iy_d\nabla_S+\tilde{y}_d\nabla_T)\nabla_\eta\,e^{i(\frac{A_1}{v_{\cal F}}+\frac{A_2}{v_g})} & 0 & 0 \\
 \frac{1-i}{2}y_d\nabla_\eta\nabla_S\,e^{i(\frac{A_1}{v_{\cal F}}+\frac{A_2}{v_g})} & y_s\nabla_\eta\,e^{-i\frac{A_2}{v_g}} & 0 \\
 3Y_d\nabla^2_S\,e^{i(2\frac{A_1}{v_{\cal F}}+3\frac{A_2}{v_g})}  & Y_s\nabla_T\nabla_S\,e^{i(\frac{A_1}{v_{\cal F}}+\frac{A_2}{v_g})} & y_b\,e^{-i\frac{A_2}{v_g}}
 \end{array}\right)}v_d\,,
 \label{Quark}
 \end{eqnarray}
where $\langle H_{u}\rangle\equiv v_{u}=v\sin\beta/\sqrt{2}$ and $\langle H_{d}\rangle\equiv v_{d}=v\cos\beta/\sqrt{2}$ with $v=246$ GeV, $v_{\cal F}=v_{\Theta}(1+\kappa^2)^{1/2}$ with $\kappa=v_S/v_\Theta$  in SUSY limit, and 
 \begin{eqnarray}
\nabla_Q\equiv \frac{v_Q}{\sqrt{2}\Lambda}\qquad\text{with}~Q=\eta, S, T, \Theta, \Psi\,.
 \end{eqnarray}
Here $\mathcal{M}_{f}= V^{f\dag}_{R}\,{\rm Diag}(m_{f_1},m_{f_2},m_{f_3})\, V^{f}_{L}$ where $f_{i}$ stands for $i$-th generation of $f$-type quark, and $V^{f}_{L}$ and $V^{f}_{R}$ are the diagonalization matrices for $\mathcal{M}^\dag_{f}\mathcal{M}_{f}$ and $\mathcal{M}_{f}\mathcal{M}^\dag_{f}$, respectively.
One of the most interesting features observed by experiments on the quarks is that the mass spectrum of the up-type quarks exhibits a much stronger hierarchical pattern to that of the down-type quarks, which may indicate that the CKM matrix\,\cite{PDG} is mainly generated by the mixing matrix of the down-type quark sector. So the following {\it new} expansion parameters could be defined in a way that the diagonalizing matrices $V^d_L$ and $V^u_L$ satisfy the CKM matrix in the Wolfenstein parametrization $V_{\rm CKM}=V^u_L\,V^{d\dag}_L$: 
\begin{eqnarray}
\nabla_T&=&\kappa\frac{|\hat{y}_d|}{|\hat{\tilde{y}}_d|}\qquad\text{with}\,\phi_{\tilde{d}}=-\phi_d-\frac{\pi}{2}\,,\label{mdi}\\
 \nabla_\Psi&\simeq&\lambda^{3/4}\Big|\frac{X_1\delta^{\rm G}_2}{X_2\delta^{\rm G}_1}\Big|^{\frac{1}{2}}\Big(\frac{B\,(1+\kappa^2)}{6\kappa^2}\frac{|\hat{y}_b|}{|\hat{Y}_d|}\Big)^{\frac{1}{4}}\,,\label{expan_0}\\
 \nabla_\Theta&=&\frac{1}{\kappa}\nabla_S=\Big|\frac{X_2\delta^{\rm G}_1}{X_1\delta^{\rm G}_2}\Big|\sqrt{\frac{2}{1+\kappa^2}}\nabla_\Psi\,,
  \label{expan_1}
\end{eqnarray}
where $\arg(\hat{y}_i)\equiv\phi_{i}$ and $B=A\sqrt{\rho^{2}+\eta^{2}}$ with the Wolfenstein parametrization\,\footnote{We take $\lambda=0.22509^{+0.00091}_{-0.00071}$, $A=0.825^{+0.020}_{-0.037}$, $\bar{\rho}=\rho/(1-\lambda^2/2)=0.160^{+0.034}_{-0.021}$, and $\bar{\eta}=\eta/(1-\lambda^2/2)=0.350^{+0.024}_{-0.024}$ with $3\sigma$ errors\,\cite{ckm}.} ($\lambda, \rho, \eta, A$)\,\cite{Wolfenstein:1983yz}. Note that the expansion parameters $\nabla_\Psi$ and $\nabla_\Theta (\nabla_S)$ associated with the $U(1)_X$ charged fields are defined by the relation Eq.\,(\ref{scale1}) associated with the two QCD anomalous $U(1)$, containing the model dependent parameter $|X_i\delta^{\rm G}_j/X_j\delta^{\rm G}_i|$ with $i\neq j$. 

From the empirical down-type quark mass ratios calculated from the measured values $(m_d/m_b)_{\rm PDG}\doteqdot1.12^{+0.13}_{-0.11}\times10^{-3}$ and $(m_s/m_b)_{\rm PDG}\doteqdot2.30^{+0.21}_{-0.12}\times10^{-2}$ with $(m_b)_{\rm PDG}\doteqdot4.18^{+0.04}_{-0.03}\,{\rm GeV}$\,\cite{PDG}, we  can obtain roughly the down-type quark mixing angles in the standard parametrization\,\cite{Chau:1984fp}
 \begin{eqnarray}
\theta^d_{12}\approx \frac{1}{\sqrt{2}}\Big|\frac{\hat{y}_d}{\hat{y}_s}\Big|\,\nabla_S\,,\qquad \theta^d_{23}\simeq \Big|\frac{\hat{Y}_s}{\hat{y}_b}\Big|\,\nabla_S\,\nabla_T^2\,,\qquad\theta^d_{13}\simeq 3\Big|\frac{\hat{Y}_d}{\hat{y}_b}\Big|\,\nabla^2_\Psi\nabla^2_S\,.
 \end{eqnarray}
And their corresponding down-type quark masses are roughly given by 
 \begin{eqnarray}
m_d\simeq2|\hat{y}_d|\,\nabla_\Psi\nabla_S\nabla_\eta|\sin\phi_d|\,v_d\,, \qquad m_s\simeq|\hat{y}_s|\,\nabla_\Psi\nabla_\eta\,v_d\,,\qquad
m_b\simeq|\hat{y}_b|\,\nabla_\Psi\,v_d\,. 
 \end{eqnarray}
Note that the parametrization of Eq.\,(\ref{mdi}) is very crucial to reproduce the $d$- and $s$-quark mass and the mixing angle $\theta^d_{12}$.
 
From the mass ratio of $t$- and $b$-quark $(m_b/m_t)_{\rm PDG}\doteqdot2.41^{+0.03}_{-0.03}\times10^{-2}$ in PDG\,\cite{PDG}
 the value of $\tan\beta\equiv v_u/v_d$ can be obtained in a good approximation:
 \begin{eqnarray}
\tan\beta\simeq\lambda^{\frac{3}{2}}\Big(\frac{m_t}{m_b}\Big)_{\rm PDG}\Big|\frac{\hat{y}_b}{\hat{y}_t}\Big|\Big|\frac{X_1\delta^{\rm G}_2}{X_2\delta^{\rm G}_1}\Big|\Big(\frac{B(1+\kappa^2)}{6\kappa^3}\Big|\frac{\hat{y}_b}{\hat{Y}_d}\Big|\Big)^{\frac{1}{2}}\,.
 \end{eqnarray}
The top Yukawa coupling $\hat{y}_t$ can be directly obtained from the top quark mass $m_t=|\hat{y}_t|v_u=173.1\pm0.6\,{\rm GeV}$\,\cite{PDG}. From the hierarchical mass ration between $u$- and $c$-quark $(m_u/m_c)_{\rm PDG}\doteqdot1.72^{+0.52}_{-0.34}\times10^{-3}$ we obtain
\begin{eqnarray}
 \Big(\frac{m_u}{m_c}\Big)_{\rm PDG}\simeq \sqrt{\frac{3}{2}}\,\Big|\frac{\hat{y}_u}{\hat{y}_c}\Big|\,\nabla^2_\Psi\nabla_S\,,
  \label{}
\end{eqnarray}
and its corresponding mixing angle
 \begin{eqnarray}
 \theta^u_{12}\simeq\frac{1}{\sqrt{2}}\,\Big|\frac{\hat{y}_u}{\hat{y}_c}\Big|\,\nabla_S\nabla^2_\Psi\,.
  \label{}
\end{eqnarray}
In turn, the expansion parameter $\nabla_\eta$ is defined by using $(m_c/m_t)_{\rm PDG}\doteqdot7.39^{+0.20}_{-0.20}\times10^{-3}$:
\begin{eqnarray}
 \nabla_\eta\simeq\lambda^{\frac{13}{4}}\Big|\frac{X_2\delta^{\rm G}_1}{X_1\delta^{\rm G}_2}\Big|^{\frac{1}{2}}\Big|\frac{\hat{y}_t}{\hat{y}_c}\Big| \Big(\frac{426\,\kappa^2}{B\,(1+\kappa^2)}\Big|\frac{\hat{Y}_d}{\hat{y}_b}\Big|\Big)^{\frac{1}{4}}\,.
  \label{}
\end{eqnarray}
As designed, with the fields redefinition the CKM matrix with $J^{\rm quark}_{CP}={\rm Im}[V_{us}V_{cb}V^{\ast}_{ub}V^{\ast}_{cs}]\simeq A^{2}\lambda^{6}\sqrt{\rho^{2}+\eta^{2}}\sin\delta^{q}_{CP}$ and its CP phase
$\delta^{q}_{CP}\equiv\phi^{d}_{2}-2\phi^{d}_{3}=\tan^{-1}\left(\eta/\rho\right)$ is well described,
 where $\phi^{d}_{2}\simeq\arg(\hat{Y}^\ast_d\hat{y}_b)-\phi^{d}_{1}/2$ and $2\phi^{d}_{3}\simeq\arg(\hat{Y}^\ast_d\hat{y}_b)+\phi^{d}_{1}-\phi^{d}_{2}$, and $\phi^{d}_{1}=\arg(\hat{Y}^\ast_s\hat{y}_b)/2$. 
 
Hence it is very crucial for obtaining the right values of the new expansion parameters to reproduce the empirical results of the CKM mixing angles and quark masses. In addition, such right values are needed to reproduce the empirical results of the charged leptons and the light active neutrino masses in our model. In the following subsequent section we will perform a numerical simulation.
%%%%%%%%%%%%%%%%%%%%%%%%%%%%%%%%%%%%%%%%%%%%%%%%%%%%%%%%%%%%%%%%%%%%%%
\subsection{Numerical analysis for Quark sector}
\label{num_quark}
We perform a numerical simulation\,\footnote{Here, in numerical calculation, we have only considered the mass matrices in Eq.\,(\ref{Quark}) since it is expected that the corrections to the VEVs due to dimensional operators contributing to Eq.\,(\ref{super_d}) could be small enough below a few percent level, see Appendix\,\ref{corre}.} using the linear algebra tools of Ref.\,\cite{Antusch:2005gp}. With the inputs 
\begin{eqnarray}
\tan\beta=7.40\,, \qquad\kappa=0.96\,,
  \label{tanpara}
\end{eqnarray}
and $|\hat{y}_d|=0.9200$ ($\phi_d=6.2100$ rad), $|\hat{\tilde{y}}_d|=3.1400$, $|\hat{y}_s|=0.3300$ ($\phi_s=2.9300$ rad), $|\hat{y}_b|=1.0100$ ($\phi_b=0$), $|\hat{y}_u|=0.3300$ ($\phi_u=0$ rad), $|\hat{y}_c|=0.4400$ ($\phi_c=5.9700$ rad), $|\hat{\tilde{y}}_c|=0.8040$ ($\phi_{\tilde{c}}=5.9900$ rad), $|\hat{y}_t|=1.0042$ ($\phi_t=0$), $|\hat{Y}_d|=2.8000$ ($\phi_{Y_d}=2.6000$ rad), $|\hat{Y}_s|=1.3200$ ($\phi_{Y_s}=5.1900$ rad), leading to 
\begin{eqnarray}
\nabla_\Psi=0.1770\,, \quad\nabla_S=0.1156\,, \quad\nabla_T=0.2813\,, \quad \nabla_\eta=0.0740\,,
 \label{quarkvalue}
 \end{eqnarray} 
we obtain the mixing angles and Dirac CP phase $\theta^q_{12}=12.9930^{\circ}$, $\theta^q_{23}=2.4339^{\circ}$, $\theta^q_{13}=0.2018^{\circ}$, $\delta^q_{CP}=64.9888^{\circ}$ compatible with the $3\sigma$ Global fit of CKMfitter\,\cite{ckm}; the masses $m_d=4.6244$ MeV, $m_s=102.8420$ MeV, $m_b=4.1682$ GeV, $m_u=2.6977$ MeV, $m_c=1.2785$ GeV, and $m_t=173.1$ GeV.  

Below the scale of spontaneous $SU(2)_L\times U(1)_Y$ gauge symmetry breaking, the running mass includes corrections from QCD and QED loops\,\cite{PDG}.
In order to explain the experimental data on quark and lepton masses\,\footnote{For charged leptons ($e,\mu,\tau$) we have used the experimental data\,\cite{PDG} in this work since the difference between pole mass and running mass are less significant.} we have used, it is meaningful to use the masses at a common momentum scale $\mu$ which is heavier than the QCD scale of about $1$ GeV.
Hence, in the $\overline{{\rm MS}}$ scheme for the light quark ($u$-, $d$-, and $s$-quark) the renomalization scale has been chosen to be a common scale $\mu\approx2$ GeV and their masses are current quark masses at $\mu\approx2$ GeV, and for heavy quarks ($b$- and $c$-quark) the renormalization scale equal to the quark mass are chosen to be $\bar{m}_Q(\mu)$ at $\mu=\bar{m}_Q$. For top quark ($t$-quark), the $t$-quark mass at scales below the pole mass is unphysical since the $t$-quark decouples at its scale, hence its mass is more directly determined by experiments, see Ref.\,\cite{PDG}, leading to the value we have used.
 
 %%%%%%%%%%%%%%%%%%%%%%%%%%%%%%%%%%%%%%%%%%%%%%%%%%%%%%%%%%%%%%%%%%%%%%
\subsection{Scale of PQ phase transition induced by Hadron sector}
\label{}
In order to obtain the QCD axion decay constant $F_A$ (or, equivalently, flavored-axion decay constants $F_{a_i}=f_{a_i}/\delta^{\rm G}_i$ through flavored-axion model\,\cite{Ahn:2016hbn}), we consider here two constraints coming from the astroparticle physics, {\it e.g.} axion cooling of stars\,\cite{WD01, Bertolami:2014wua, Leinson:2014cja, Leinson:2014ioa, Sedrakian:2015krq}, and flavor-violating processes induced by the flavored-axions, {\it e.g.} $K^+\rightarrow \pi^++A_i$, etc.\,\cite{Artamonov:2008qb, Wilczek:1982rv, Adler:2008zza, Fantechi:2014hqa, flavoraxion}.

\noindent{\bf(i)} Below the chiral symmetry breaking scale, the axion-hadron interactions are meaningful for the axion production rate in the core of a star where the temperature is not as high as 1 GeV, which is given by
 \begin{eqnarray}
  -{\cal L}^{a-\psi_N} &=& \frac{\partial_{\mu}a}{2F_{A}}\,X_{\psi_N}\overline{\psi}_N\,\gamma_\mu\gamma^5\,\psi_N
  \label{a_nucleon}
 \end{eqnarray}
where the QCD axion decay constant is given by $F_A=f_A/N$ with $f_A=\sqrt{2}\,\delta^{\rm G}_2f_{a_1}=\sqrt{2}\,\delta^{\rm G}_1f_{a_2}$, and $\psi_N$ is the nucleon doublet $(p,n)^T$ (here $p$ and $n$ correspond to the proton field and neutron field, respectively). The couplings of the axion to the nucleon can be rewritten as\,\cite{Ahn:2016hbn}
 \begin{eqnarray}
  -{\cal L}_{A} &\supset&\frac{\partial^{\mu}a}{2F_A}\Big\{\Big(\frac{\tilde{X}_u}{N}-\frac{1}{1+z+\omega}\Big)\bar{u}\gamma^{\mu}\gamma_5u+\Big(\frac{\tilde{X}_d}{N}-\frac{z}{1+z+\omega}\Big)\bar{d}\gamma^{\mu}\gamma_5d\nonumber\\
  &&+\Big(\frac{\tilde{X}_s}{N}-\frac{\omega}{1+z+\omega}\Big)\bar{s}\gamma^{\mu}\gamma_5s\Big\}\,,
   \label{Nucleon_Lagran01}
 \end{eqnarray}
 where $\tilde{X}_q=\delta^{\rm G}_{2}X_{1q}+\delta^{\rm G}_{1}X_{2q}$ with $q=u,d,s$ and $X_{1u}=8$, $X_{1d}=8$, $X_{1s}=0$, $X_{2u}=3$, $X_{2d}=1$, $X_{2s}=-1$. 
 From Eqs.\,(\ref{a_nucleon}-\ref{Nucleon_Lagran01}) the QCD axion coupling to the neutron can be obtained as
 \begin{eqnarray}
  g_{Ann}=\frac{X_n\,m_n}{F_A}\,,
  \label{coupling_n01}
 \end{eqnarray}
where the neutron mass $m_n=939.6$ MeV, and the axion-neutron coupling, $X_{n}$, related to axial-vector current matrix elements by Goldberger-Treiman relations\,\cite{PDG} is obtained as
 \begin{eqnarray}
  X_{n}&=&%\Big(\frac{4}{\delta^{\rm G}_1}+\frac{3}{2\delta^{\rm G}_2}-\eta\Big)\Delta d+\Big(\frac{4}{\delta^{\rm G}_1}+\frac{1}{2\delta^{\rm G}_2}-\eta z\Big)\Delta u-\Big(\frac{1}{2\delta^{\rm G}_2}+\eta\omega\Big)\Delta s\,,
  \Big(\frac{3}{4}-\eta\Big)\Delta d+\Big(\frac{5}{12}-\eta z\Big)\Delta u-\Big(\frac{1}{6}+\eta\omega\Big)\Delta s\,,
  \label{coupling_n02}
 \end{eqnarray}
where $\eta=(1+z+\omega)^{-1}$ with $z=m_u/m_d$ and $\omega=m_u/m_s\ll z$, and the $\Delta q$ are given by the axial vector current matrix element $\Delta q\,S_\mu=\langle p|\bar{q}\gamma_\mu\gamma^5q|p\rangle$. 
 Now, for numerical estimations on Eq.\,(\ref{coupling_n01}) we adopt the central values of $\Delta u=0.84\pm0.02$, $\Delta d=-0.43\pm0.02$ and $\Delta s=-0.09\pm0.02$, and take the Weinberg value for $0.38<z<0.58$\,\cite{PDG} and $\omega=0.315\,z$. 
Then, the value of the axion-neutron coupling lies in ranges $0.007\lesssim X_n\lesssim0.111$.
There is a hint for extra cooling from the neutron star in the supernova remnant ``Cassiopeia A" by axion neutron bremsstrahlung, requiring a coupling to the neutron of size $g_{Ann}=(3.8\pm3)\times10^{-10}$\,\cite{Leinson:2014ioa}, which is translated into $9.94\times10^{6}\lesssim F_A/{\rm GeV}\lesssim1.31\times10^{9}$. However, since the cooling of the superfluid core in the neutron star can also be explained by neutrino emission in pair formation in a multicomponent superfluid state $^3{\rm P}_2(m_j=0,\pm1,\pm2)$\,\cite{Leinson:2014cja}, one may not take it seriously. The range quoted is compatible with the state-of-the-art upper limit on the coupling from neutron star cooling $g_{Ann}<8\times10^{-10}$\,\cite{Sedrakian:2015krq}, whose upper bound is interpreted as the lower bound of the QCD axion decay constant:
 \begin{eqnarray}
   F_A>(0.84-13.08)\times10^{7}\,{\rm GeV}\,.
  \label{a_nucleon01}
 \end{eqnarray}

\noindent{\bf(ii)} Since a direct interaction of the SM gauge singlet flavon fields charged under $U(1)_X$ %responsible for spontaneous symmetry breaking 
with the SM quarks charged under $U(1)_X$ can arise through Yukawa interaction, the flavored-axion interactions with the flavor violating coupling to the $s$- and $d$-quark is given by
\begin{eqnarray}
 {\cal L}^{A_isd}_Y\simeq i\Big(\frac{|X_1|\,A_1}{2f_{a_1}}+\frac{|X_2|\,A_2}{f_{a_2}}\Big)\bar{s}d\,(m_s-m_d)\lambda\Big(1-\frac{\lambda^2}{2}\Big)\,,
  \label{}
\end{eqnarray}
where\,\footnote{Actually, in the standard parametrization the mixing elements of $V^d_R$ are given by $\theta^R_{23}\simeq A\lambda^2\nabla_\eta\,|\hat{y}_s/\hat{y}_b|$, $\theta^R_{13}\simeq\sqrt{2}\,B\lambda^3\,\nabla_\eta\nabla_S$, and $\theta^R_{12}\simeq\sqrt{2}|\hat{y}_d/\hat{y}_s|^2\cos\phi_{\tilde{d}}\,\nabla^2_S$. Its effect to the flavor violating coupling to the $s$- and $d$-quark is negligible: $(V^d_R\,{\rm Diag.}(\frac{A_1}{v_{\cal F}}+\frac{2A_2}{v_{g}}, \frac{A_1}{v_{\cal F}}+\frac{2A_2}{v_{g}},0)\,V^{d\dag}_R)_{12}=0$ at leading order.} $V^{d\dag}_L\approx V_{\rm CKM}$, $f_{a_1}=|X_1|v_{\cal F}$, and $f_{a_2}=|X_2|v_g$ are used. Then the decay width of $K^+\rightarrow\pi^++A_i$ is given by\,\cite{Wilczek:1982rv, Feng:1997tn, Ahn:2018cau}
 \begin{eqnarray}
   \Gamma(K^+\rightarrow\pi^++A_i)=\frac{m^3_K}{16\pi}\Big(1-\frac{m^2_{\pi}}{m^2_{K}}\Big)^3\big|{\cal M}_{dsi}\big|^2\,,
  \label{}
 \end{eqnarray}
 where $m_{K^{\pm}}=493.677\pm0.013$ MeV, $m_{\pi^{\pm}}=139.57018(35)$ MeV\,\cite{PDG}, and
 \begin{eqnarray}
   \big|{\cal M}_{ds1}\big|^2=\Big|\frac{1}{f_{a_1}/k_2}\lambda\Big(-1+\frac{\lambda^2}{2}\Big)\Big|^2\,,\qquad \big|{\cal M}_{ds2}\big|^2=\Big|\frac{1}{f_{a_2}/k_1}\lambda\Big(-1+\frac{\lambda^2}{2}\Big)\Big|^2\,.
  \label{}
 \end{eqnarray}
From the present experimental upper bound ${\rm Br}(K^+\rightarrow\pi^+A_i)<7.3\times10^{-11}$\,\cite{Adler:2008zza} with ${\rm Br}(K^+\rightarrow\pi^+\nu\bar{\nu})=1.73^{+1.15}_{-1.05}\times10^{-10}$\,\cite{Artamonov:2008qb},
we obtain the lower limits of flavored-axion decay constants and their corresponding QCD axion decay constant
 \begin{eqnarray}
     \begin{array}{ll}
              f_{a1}>|k_2|\times1.15\times10^{11}\,{\rm GeV}  \\
              f_{a2}>|k_1|\times1.15\times10^{11}\,{\rm GeV}
             \end{array}
     \,~\Leftrightarrow~\,\begin{array}{ll}
              F_{A}=\frac{f_{a_1}}{4\,|k_2|\sqrt{2}}>2.03\times10^{10}\,{\rm GeV}  \\
              F_{A}=\frac{f_{a_2}}{3\,|k_1|\sqrt{2}}>2.72\times10^{10}\,{\rm GeV}
             \end{array}\,,
 \label{cons_2}
 \end{eqnarray}
 where $F_A=f_{a_i}/(\delta^{\rm G}_i\sqrt{2})$ is used.
% \begin{eqnarray}
%  {\rm Br}(K^+\rightarrow\pi^+A_i)&=&\frac{\Gamma(K^+\rightarrow\pi^+A_i)}{\Gamma(K^+\rightarrow\pi^+\nu\bar{\nu})}\,{\rm Br}(K^+\rightarrow\pi^+\nu
%\bar{\nu})\nonumber\\
%  &\simeq&\frac{\Gamma(K^+\rightarrow\pi^+A_i)}{8.47\times10^{15}\,{\rm GeV}}\,,
% \end{eqnarray}
Note that the lower bounds of flavored-axion decay constants $f_{a_i}$ are dependent on the values of $k_i$, while the QCD axion decay constant $F_A$ does depend on the properties ($2\alpha$ and $\omega$ in Eq.\,(\ref{dwahn})) from the QCD instanton background instead of the $k_i$. %Clearly, from Eqs.\,(\ref{a_nucleon01}) and (\ref{cons_2}) we see that the strongest lower bound of QCD axion decay constant comes from the present upper limit of ${\rm Br}(K^+\rightarrow\pi^+A_i)<7.3\times10^{-11}$\,\cite{Adler:2008zza}.
Clearly, from Eqs.\,(\ref{a_nucleon01}) and (\ref{cons_2}) the most stringent constraint on the QCD axion decay constant comes from the present experimental upper bound ${\rm Br}(K^+\rightarrow\pi^+A_i)<7.3\times10^{-11}$\,\cite{Adler:2008zza}
 \begin{eqnarray}
  F_{A}>2.72\times10^{10}\,{\rm GeV}\,.
  \label{k_bound}
 \end{eqnarray}
In the near future the NA62 experiment will be expected to reach the sensitivity of ${\rm Br}(K^+\rightarrow\pi^++A_i)<1.0\times10^{-12}$\,\cite{Fantechi:2014hqa}, which is interpreted as the flavored-axion decay constant and its corresponding QCD axion decay constant 
 \begin{eqnarray}
 f_{a_i}>9.86\times10^{11}\,{\rm GeV}\,\Leftrightarrow\, F_{A}>2.32\times10^{11}\,{\rm GeV}\,.
  \label{knew_bound}
 \end{eqnarray}

 %%%%%%%%%%%%%%%%%%%%%%%%%%%%%%%%%%%%%%%%%%%%%%%%%%%%%%%%%%%%%%%%%%%%%%%%%%%%
\section{Leptons and flavored-Axions}

Next, %according to the $\mu$--$\tau$ power law in Ref.\,\cite{Ahn:2016hbn, Ahn:2014gva},  
we assign the left-handed charged lepton $SU(2)_{L}$ doublets denoted as $L_e,\,L_\mu,\,L_\tau$  to the $({\bf 1},  -p-{\cal Q}_{y^\nu_1})$, $({\bf 1}',  -p-{\cal Q}_{y^\nu_1})$, and $({\bf 1}'', -p-{\cal Q}_{y^\nu_1})$, respectively, while the right-handed charged leptons denoted as $e^c,\,\mu^c$ and $\tau^c$, the electron flavor to the $({\bf 1}, p+{\cal Q}_{y^\nu_1}+6q)$, the muon flavor to the $({\bf 1}'', p+{\cal Q}_{y^\nu_1}-3q)$, and the tau flavor to the $({\bf 1}', p+{\cal Q}_{y^\nu_1}-q)$. And we assign the right-handed neutrinos $SU(2)_L$ singlets denoted as $N^c$ to the $({\bf 3}, p)$. Note that ${\cal Q}_{y^\nu_1}={\cal Q}_{y^\nu_2}={\cal Q}_{y^\nu_3}$ is assigned to give a tribimaximal (TBM)-like mixing pattern.
In addition, additional Majorana fermions are introduced to have no axionic domain-wall problem, which link low energy neutrino oscillations to astronomical-scale baseline neutrino oscillations.
Under $SL_2(F_3)\times U(1)_X$ we assign the additional Majorana neutrinos $SU(2)_L$ singlets denoted as $S^c_e$, $S^c_\mu$ and $S^c_\tau$ to the $({\bf 1}, p+{\cal Q}_{y^\nu_1}-{\cal Q}_{y^s_1})$, $({\bf 1}'', p+{\cal Q}_{y^\nu_1}-{\cal Q}_{y^s_2})$ and $({\bf 1}', p+{\cal Q}_{y^\nu_1}-{\cal Q}_{y^s_3})$, respectively.  Here ${\cal Q}_{\cal Y}$ denotes the $U(1)_X$ quantum number of Yukawa coupling ${\cal Y}$ which appears in the superpotentials (\ref{lagrangian_q}) and (\ref{lagrangian2}) sewed by the five (among seven) in-equivalent representations ${\bf 1}$, ${\bf 1}'$, ${\bf 1}''$, ${\bf 2}'$ and ${\bf 3}$ of $SL_2(F_3)$.%: for example, ${\cal Q}_{y_u}=-3q$, ${\cal Q}_{y_c}=-q$, ${\cal Q}_{Y_d}=-3q$, ${\cal Q}_{Y_s}=-q$, ${\cal Q}_{y_d}=-q$, ${\cal Q}_{y_s}=q$, ${\cal Q}_{y_b}=q$.%${\cal Q}_{y_\tau}=q$, ${\cal Q}_{y_\mu}=3q$, ${\cal Q}_{y_e}=-6q$. 

%In the previous section, we have shown that the SM quark masses and mixings could well be described by the new expansion parameters defined under the $U(1)_X\times[gravity]^2$ anomaly-free condition. Along this line of quark sector, in this section we describe the Yukawa superpotential for leptons and flavored-axions. In turn, we investigate how the $U(1)_X$ quantum number of leptons are well flavor-structured and such quantum numbers can provide predictions on the neutrino mass ordering, $\delta_{CP}$ and $\theta_{23}$. 

As mentioned before, with the conditions (\ref{cond}) and (\ref{ux_gr}) satisfied, new additional Majorana fermions $S^{c}_{e,\,\mu,\,\tau}$ besides the heavy Majorana neutrinos are introduced in the lepton sector. Hence, such new additional Majorana neutrinos can play a role of the active neutrinos as pseudo-Dirac neutrinos.
Under $SL_2(F_3)\times U(1)_{X}$ with $U(1)_R=+1$, the quantum numbers of the lepton fields are summarized as in Table\,\ref{reps_l}.
%\begin{center}
\begin{table}[h]
\caption{\label{reps_l} Representations of the lepton fields under $SL_2(F_3)\times U(1)_{X}$ with $U(1)_R=+1$. And here $r\equiv Q_{y^{\nu}_1}+p$ is defined.}
\begin{ruledtabular}
\begin{tabular}{ccccc}
Field &$L_{e},L_{\mu},L_{\tau}$&$e^c,\mu^c,\tau^c$&$N^{c}$&$S_e^c,S_\mu^c,S_\tau^c$\\
\hline
$SL_2(F_3)$&$\mathbf{1}$, $\mathbf{1^{\prime}}$, $\mathbf{1^{\prime\prime}}$&$\mathbf{1}$, $\mathbf{1^{\prime\prime}}$, $\mathbf{1^\prime}$&$\mathbf{3}$&$\mathbf{1}$, $\mathbf{1^{\prime\prime}}$, $\mathbf{1^{\prime}}$\\
$U(1)_{X}$& $ -r $ & $r-{\cal Q}_{y_e}, r-{\cal Q}_{y_\mu}, r-{\cal Q}_{y_\tau}$& $p$&$r-{\cal Q}_{y^s_1}$, $r-{\cal Q}_{y^s_2}$, $r-{\cal Q}_{y^s_3}$\\
%$U(1)_R$& $ 1 $ & $1$~~~& $1$& $1$\\
%$SU(2)\times U(1)_Y$&$2_{-1}$&$2_\frac{1}{3}$&$1_{\frac{4}{3}}$&$1_{-\frac{2}{3}}$&$1_{-2}$&$1_{0}$&$2_{1}$&$2_{-1}$\\
\end{tabular}
\end{ruledtabular}
\end{table}
%\end{center} 
The lepton Yukawa superpotential, similar to the quark sector, invariant under $G_{\rm SM}\times G_F\times U(1)_R$ reads
%\begin{widetext}
\begin{eqnarray}
 W_{\ell\nu} &=&
y_\tau\,\tau^c L_\tau H_d+y_\mu\,\mu^c L_\mu H_d+y_e\,e^c L_e H_d\nonumber\\
 &+&y^s_1\,S^c_eL_eH_u+y^s_2\, S^c_\mu L_\mu H_u+y^s_3\,S^c_\tau L_\tau  H_u\nonumber\\
 &+& \big\{y^{\nu}_{1}(N^c\Phi_T)_{{\bf 1}}L_{e}+y^{\nu}_{2}(N^c\Phi_T)_{{\bf 1}''}L_{\mu}+y^{\nu}_{3}(N^c\Phi_T)_{{\bf 1}'}L_{\tau}\big\}\frac{H_u}{\Lambda}\nonumber\\
 &+&\frac{1}{2}(\hat{y}_\Theta\Theta+\hat{y}_{\tilde{\Theta}}\tilde{\Theta})(N^{c}N^{c})_{{\bf 1}}+\frac{\hat{y}_R}{2}(N^{c}N^{c})_{{\bf 3}} \Phi_S\nonumber\\  
 &+&\frac{1}{2}\{y^{ss}_1\,S^c_eS^c_e+y^{ss}_2\,S^c_\mu S^c_\tau+y^{ss}_2\,S^c_\tau S^c_\mu\}\Theta\,.
 \label{lagrangian2}
 \end{eqnarray}
%\end{widetext}
Remark that, as in the SM quark fields since the $U(1)_X$ quantum numbers are arranged to lepton fields as in Table\,\ref{reps_l} with the conditions (\ref{cond}) and (\ref{ux_gr}) satisfied,  it is expected that the SM gauge singlet flavon fields derive higher-dimensional operators, which are eventually visualized into the Yukawa couplings of leptons as a function of flavon fields $\Psi(\tilde{\Psi})$.

For pseudo-Dirac neutrino as the active neutrino to be realized in a way that the neutrino oscillations at low energies could have a direct connection to new neutrino oscillations available on high-energy neutrinos\,\cite{Ahn:2016hbn}, two requirements are needed since the quantum numbers $L_{e\,,\mu\,,\tau}$ (or equivalently $Q_{y^{\nu}_i}$) are not uniquely determined: (i) the quantum numbers ${\cal Q}_{y^\nu_i}$ and ${\cal Q}_{y^s_i}$ should have opposite sign due to ${\cal Q}_{y^{ss}_1}=2({\cal Q}_{y^s_1}-{\cal Q}_{y^\nu_1})$ and ${\cal Q}_{y^{ss}_2}={\cal Q}_{y^{ss}_3}={\cal Q}_{y^s_2}+{\cal Q}_{y^s_3}-2{\cal Q}_{y^\nu_1}$, (ii) especially, the quantum numbers ${\cal Q}_{y^s_2}$ and ${\cal Q}_{y^s_3}$ should have the same sign for normal neutrino mass ordering, and (iii)
\begin{eqnarray}
|{\cal Q}_{y^{ss}_i}|\gg|{\cal Q}_{y^{s}_i}|\gg|{\cal Q}_{y^{\nu}_i}|\,,
\label{cond_pseudo}
 \end{eqnarray}
As we shall see later, it could make a connection between the neutrino oscillation at low energies  and new oscillations available on high-energy neutrinos through astronomical-scale baseline. % characterized by $\Delta m^2_k=2|\delta^\nu_k|m_{\nu_k}$. 
Then,
the quantum numbers ${\cal Q}_{y^s_i}$ can be uniquely determined by taking into account both the $U(1)_X\times[gravity]^2$ anomaly-free condition in Eq.\,(\ref{cond1}) and the hat Yukawa coupling of order unity, $1/\sqrt{10}\lesssim|\hat{y}^s_i|\lesssim\sqrt{10}$, we obtain
(i) $|{\cal Q}_{y^s_3}|\gg|{\cal Q}_{y^s_1}|\geq|{\cal Q}_{y^s_2}|$ for inverted mass ordering (IO), and (ii) $|{\cal Q}_{y^s_1}|\gg|{\cal Q}_{y^s_2}|\geq|{\cal Q}_{y^s_3}|$ for normal mass ordering (NO). In such case, considering the observed neutrino mass hierarchy $\Delta m^2_{\rm sol}\equiv m^2_{\nu_2}-m^2_{\nu_1}\simeq7.50\times10^{-5}\,{\rm eV}^2$ and $\Delta m^2_{\rm atm}\simeq2.52\times10^{-3}\,{\rm eV}^2$ where $\Delta m^2_{\rm atm}\equiv m^2_{\nu_3}-m^2_{\nu_1}$ for NO; $|m^2_{\nu_2}-m^2_{\nu_3}|$ for IO, we have the followings: 
\begin{description}
\item[For the {\it case-I} with $E/N=3.83$ in Eq.\,(\ref{cas})] the Yukawa couplings of charged-leptons are represented with ${\cal Q}_{y_\tau}=-q$, ${\cal Q}_{y_\mu}=3q$, ${\cal Q}_{y_e}=-6q$ as
\begin{eqnarray}
y_e=\hat{y}_e\Big(\frac{\Psi}{\Lambda}\Big)^6\,, \qquad y_\mu=\hat{y}_\mu\Big(\frac{\tilde{\Psi}}{\Lambda}\Big)^3\,,\qquad y_\tau=\hat{y}_\tau\Big(\frac{\Psi}{\Lambda}\Big)\,;
 \label{caseI_lep}
 \end{eqnarray}  
the $U(1)_X$ quantum numbers of Yukawa couplings of pseudo-Dirac neutrinos are given for $k_1=+k_2=1$ in Eq.\,(\ref{k_val}) as 
 \begin{eqnarray}
     \begin{array}{ll}
              {\cal Q}_{y^s_1}=63q\,, ~~\qquad {\cal Q}_{y^s_2}=-18q\,,\qquad {\cal Q}_{y^{s}_3}=-17q  &
             \,;\quad\text{NO}\\
              {\cal Q}_{y^s_1}=\mp17q\,, \qquad {\cal Q}_{y^s_2}=\pm17q\,,\qquad {\cal Q}_{y^{s}_3}=28q &\,;\quad\text{\,IO}
             \end{array}\,.
 \label{caseI}
 \end{eqnarray}
Here for NO the quantum numbers ${\cal Q}_{y^s_2}$ and ${\cal Q}_{y^s_3}$ should have the same sign, while for IO ${\cal Q}_{y^s_1}$ and ${\cal Q}_{y^s_2}$ should have the opposite sign.
\item[For the {\it case-II} with $E/N=3.16$ in Eq.\,(\ref{cas})] the Yukawa couplings of charged-leptons are represented with ${\cal Q}_{y_\tau}=q$, ${\cal Q}_{y_\mu}=3q$, ${\cal Q}_{y_e}=-6q$ as
\begin{eqnarray}
y_e=\hat{y}_e\Big(\frac{\Psi}{\Lambda}\Big)^6\,, \qquad y_\mu=\hat{y}_\mu\Big(\frac{\tilde{\Psi}}{\Lambda}\Big)^3\,,\qquad y_\tau=\hat{y}_\tau\Big(\frac{\tilde{\Psi}}{\Lambda}\Big)\,;
 \end{eqnarray}  
the $U(1)_X$ quantum numbers of Yukawa couplings of pseudo-Dirac neutrinos are given for $k_1=+k_2=1$ in Eq.\,(\ref{k_val}) as 
 \begin{eqnarray}
     \begin{array}{ll}
              {\cal Q}_{y^s_1}=61q\,, ~~\qquad {\cal Q}_{y^s_2}=-18q\,,\qquad {\cal Q}_{y^{s}_3}=-17q &
             \,;\qquad\text{NO}\\
              {\cal Q}_{y^s_1}=\mp17q\,, \qquad {\cal Q}_{y^s_2}=\pm17q\,,\qquad {\cal Q}_{y^{s}_3}=26q & \,;\qquad\text{~IO}
     \end{array}\,.
 \label{caseII}
 \end{eqnarray}
\item[For the {\it case-III} with $E/N=1.83$ in Eq.\,(\ref{cas})]  the Yukawa couplings of charged-leptons are represented with ${\cal Q}_{y_\tau}=-q$, ${\cal Q}_{y_\mu}=-3q$, ${\cal Q}_{y_e}=6q$ as
\begin{eqnarray}
y_e=\hat{y}_e\Big(\frac{\tilde{\Psi}}{\Lambda}\Big)^6\,, \qquad y_\mu=\hat{y}_\mu\Big(\frac{\Psi}{\Lambda}\Big)^3\,,\qquad y_\tau=\hat{y}_\tau\Big(\frac{\Psi}{\Lambda}\Big)\,;
 \end{eqnarray}  
the $U(1)_X$ quantum numbers of Yukawa couplings of pseudo-Dirac neutrinos are given for $k_1=+k_2=1$ in Eq.\,(\ref{k_val}) as 
 \begin{eqnarray}
     \begin{array}{ll}
              {\cal Q}_{y^s_1}=57q\,, ~~\qquad {\cal Q}_{y^s_2}=-18q\,,\qquad {\cal Q}_{y^{s}_3}=-17q  &
             \,;\qquad\text{NO}\\
              {\cal Q}_{y^s_1}=\mp17q\,, \qquad {\cal Q}_{y^s_2}=\pm17q\,,\qquad {\cal Q}_{y^{s}_3}=22q  & \,;\qquad\text{~IO}
             \end{array}\,.
 \label{caseIII}
 \end{eqnarray}
\end{description}
The hat Yukawa couplings $\hat{y}_{e,\mu,\tau}$ are fixed by the numerical values in Eq.\,(\ref{quarkvalue}) used in quark sector via the empirical ratios
 $m_e/m_\mu\doteqdot4.84\times10^{-3}$, $m_\mu/m_\tau\doteqdot5.95\times10^{-2}$, and $m_\tau/m_b\doteqdot0.43$ in\,\cite{PDG} as 
 \begin{eqnarray}
\hat{y}_e=0.713\,, \qquad\hat{y}_\mu=0.818\,, \qquad\hat{y}_\tau=0.431\,.
   \label{}
\end{eqnarray}
Through the $U(1)_X$ quantum numbers of Yukawa couplings of pseudo-Dirac neutrino sector, ${\cal Q}_{y^s_i}$ ($i=1,2,3$), as shown in Eqs.\,(\ref{caseI}), (\ref{caseII}) and (\ref{caseIII}), the active neutrino mass spectra can be determined in terms of the new expansion parameters in Eq.\,(\ref{expan_1}) defined in quark sector; for example, in {\it case-I}, for NO (${\cal Q}_{y^s_1}=63q$, ${\cal Q}_{y^s_2}=-18q$, ${\cal Q}_{y^{s}_3}=-17q$):
 \begin{eqnarray}
  m_{\nu_1}&\simeq&\hat{y}^s_1\,\lambda^{\frac{189}{4}}\Big|\frac{X_1\delta^{\rm G}_2}{X_2\delta^{\rm G}_1}\Big|^{\frac{63}{2}}\Big(\frac{B\,(1+\kappa^2)}{6\kappa^2}\frac{|\hat{y}_b|}{|\hat{Y}_d|}\Big)^{\frac{63}{4}}v_u\,,\nonumber\\
 m_{\nu_2}&\simeq&\hat{y}^s_2\,\lambda^{\frac{27}{2}}\Big|\frac{X_1\delta^{\rm G}_2}{X_2\delta^{\rm G}_1}\Big|^{9}\Big(\frac{B\,(1+\kappa^2)}{6\kappa^2}\frac{|\hat{y}_b|}{|\hat{Y}_d|}\Big)^{\frac{9}{2}}v_u\,,\nonumber\\
 m_{\nu_3}&\simeq&\hat{y}^s_3\,\lambda^{\frac{51}{4}}\Big|\frac{X_1\delta^{\rm G}_2}{X_2\delta^{\rm G}_1}\Big|^{\frac{17}{2}}\Big(\frac{B\,(1+\kappa^2)}{6\kappa^2}\frac{|\hat{y}_b|}{|\hat{Y}_d|}\Big)^{\frac{17}{4}}v_u\,,
 \label{nu_mass_cn1}
 \end{eqnarray}
and, with the value $\nabla_\Psi$ in Eq.\,(\ref{quarkvalue}) obtained in quark sector the neutrino parameters are fixed within the $3\sigma$ constraints of the low energy neutrino oscillations\,\cite{Esteban:2016qun} as 
\begin{eqnarray}
\hat{y}^s_{2}\ni(1.67, 1.79)\,,\qquad \hat{y}^s_{3}\ni(1.73, 1.82)\,\qquad \hat{y}^s_{1}={\cal O}(1)\,;
  \label{nu_n1}
\end{eqnarray}
for IO (${\cal Q}_{y^s_1}=\mp17q$, ${\cal Q}_{y^s_2}=\pm17q$, ${\cal Q}_{y^{s}_3}=28q$):
 \begin{eqnarray}
  m_{\nu_1}&\simeq&\hat{y}^s_1\,\lambda^{\frac{51}{4}}\Big|\frac{X_1\delta^{\rm G}_2}{X_2\delta^{\rm G}_1}\Big|^{\frac{17}{2}}\Big(\frac{B\,(1+\kappa^2)}{6\kappa^2}\frac{|\hat{y}_b|}{|\hat{Y}_d|}\Big)^{\frac{17}{4}}v_u\,,\nonumber\\
 m_{\nu_2}&\simeq&\hat{y}^s_2\,\lambda^{\frac{51}{4}}\Big|\frac{X_1\delta^{\rm G}_2}{X_2\delta^{\rm G}_1}\Big|^{\frac{17}{2}}\Big(\frac{B\,(1+\kappa^2)}{6\kappa^2}\frac{|\hat{y}_b|}{|\hat{Y}_d|}\Big)^{\frac{17}{4}}v_u\,,\nonumber\\
 m_{\nu_3}&\simeq&\hat{y}^s_3\,\lambda^{21}\Big|\frac{X_1\delta^{\rm G}_2}{X_2\delta^{\rm G}_1}\Big|^{14}\Big(\frac{B\,(1+\kappa^2)}{6\kappa^2}\frac{|\hat{y}_b|}{|\hat{Y}_d|}\Big)^{7}v_u\,.
 \label{nu_mass_ci1}
 \end{eqnarray}
and within the $3\sigma$ constraints of the low energy neutrino oscillations\,\cite{Esteban:2016qun} by using the value $\nabla_\Psi$ in Eq.\,(\ref{quarkvalue})
\begin{eqnarray}
\hat{y}^s_{1,2}\ni(1.70, 1.81)\,,\qquad \hat{y}^s_{3}={\cal O}(1)\,.
  \label{nu_i1}
\end{eqnarray}
However, there still remain two physical parameters undetermined, the scale of $U(1)_X$ symmetry breakdown and ${\cal Q}_{y^\nu_i}$, which correspond to the physical observables, the QCD axion mass and mass splittings $\Delta m^2_k$ for new neutrino oscillations through astronomical-scale baseline. Note that the neutrino mixing angles can be determined through the lepton Yukawa superpotential in Eq.\,(\ref{lagrangian2}) structured by the $SL_2(F_3)$ symmetry together with the desired VEV directions in Eqs.\,(\ref{vevdirection1}, \ref{vevdirection2}, \ref{vevdirection3}), as will be seen later.

 %%%%%%%%%%%%%%%%%%%%%%%%%%%%%%%%%%%%%%%%%%%%%%%%%%%%%%%%%%%%%%%%%%%%%%%%%%%%
\subsection{Scale of PQ phase transition induced by Lepton sector}
Now we are going to try to fix the scale of $U(1)_X$ symmetry breakdown, together with the constraints coming from the previous quark sector, by taking flavored-axion $A_2$ coupling to electron coming from the axion cooling of stars into account.
Once the scale $f_{a2}=|X_2|\sqrt{2}\,v_{\Psi}$ is constrained by the constraints coming from rare flavor violating decay processes induced by flavored axions and axion cooling of stars, the scale $f_{a1}=|X_1|\sqrt{1+\kappa^2}\,v_\Theta$ associated to the seesaw scale could automatically be determined through Eq.\,(\ref{scale1}). 

\begin{figure}[h]
%\vspace*{-5.0cm}
%\hspace*{-1cm}
\begin{minipage}[h]{7.4cm}
\epsfig{figure=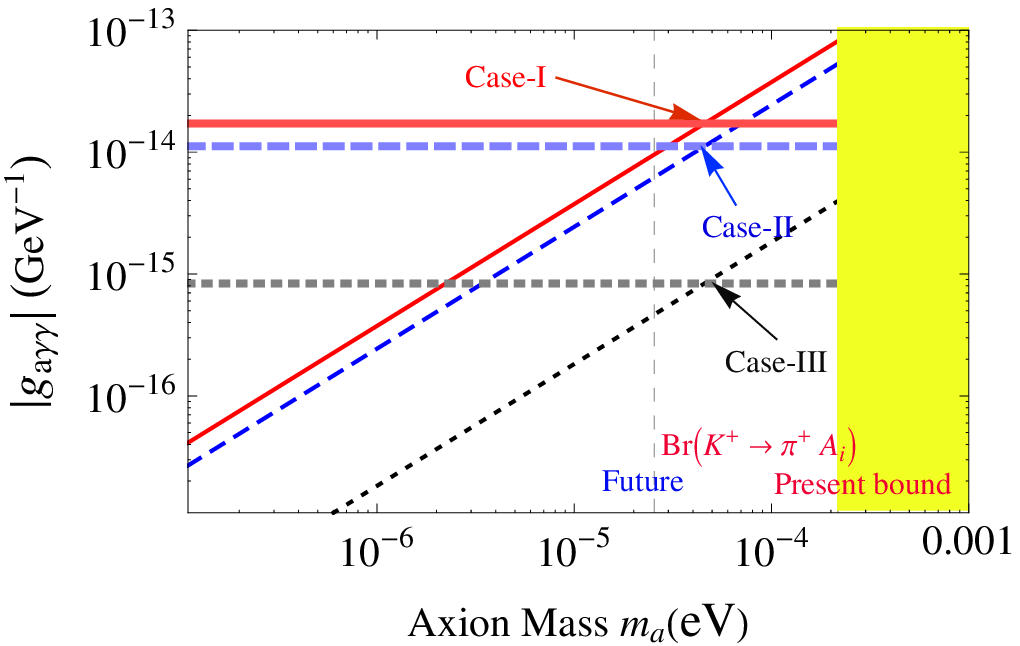,width=7.8cm,angle=0}
\end{minipage}
\hspace*{1.0cm}
\begin{minipage}[h]{7.4cm}
\epsfig{figure=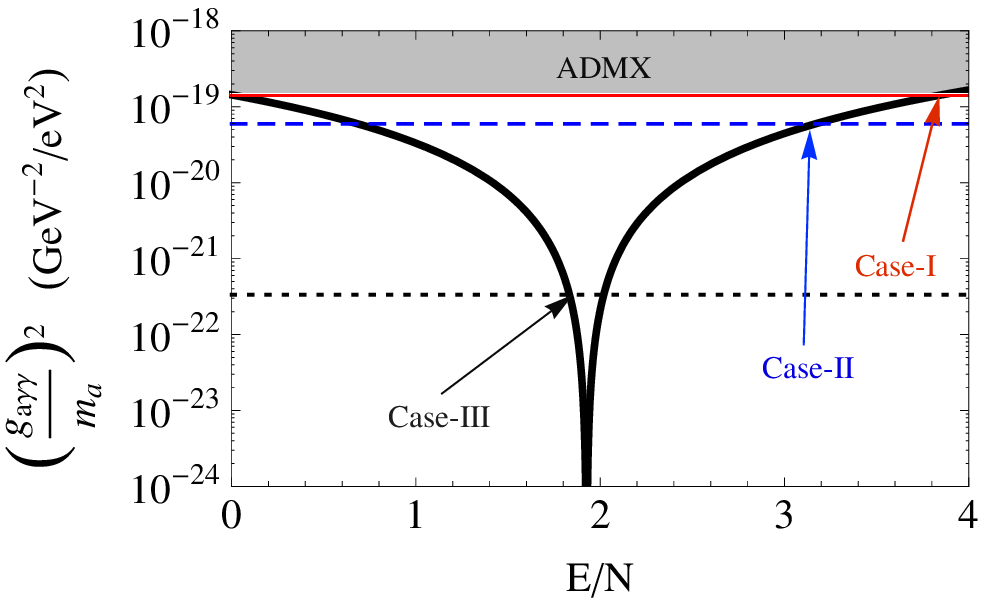,width=7.8cm,angle=0}
\end{minipage}
%\vspace*{-5.5cm}
\caption{\label{Fig1} Left plot for axion photon coupling $|g_{a\gamma\gamma}|$ as a function of the QCD axion mass $m_{a}$. Horizontal solid-red, dashed-blue, and dotted-black lines crossed by solid-red (case-I), dotted-blue (case-II), and dashed-black (case-III) lines show the model predictions for $F_A=1.29\times10^{11}$ GeV expected from the future sensitivity of ${\rm Br}(K^+\rightarrow\pi^+A_i)$: $|g_{a\gamma\gamma}|=1.72\times10^{-14}\,{\rm GeV}^{-1}$, $1.12\times10^{-14}\,{\rm GeV}^{-1}$, and $8.37\times10^{-16}\,{\rm GeV}^{-1}$, respectively, with $m_a=45.8\,\mu\,{\rm eV}$. The yellow-band indicates the excluded region derived from the present bound on ${\rm Br}(K^+\rightarrow\pi^+A_i)<7.3\times10^{-11}$\,\cite{Adler:2008zza} (equivalently $m_a<217\,\mu$eV), while the vertical black-dashed line stands for the NA62 experiment future expected sensitivity of ${\rm Br}(K^+\rightarrow\pi^+A_i)<1.0\times10^{-12}$\,\cite{Fantechi:2014hqa} (equivalently $m_a<25.5\,\mu$eV). Right plot of $(g_{a\gamma\gamma}/m_a)^2$ versus $E/N$ for $z=0.56$. The gray-band represents the experimentally excluded  bound $(g_{a\gamma\gamma}/m_{a})^2\leq1.44\times10^{-19}\,{\rm GeV}^{-2}\,{\rm eV}^{-2}$ from ADMX\,\cite{Asztalos:2003px}. Here the solid-red, dashed-blue, and dotted-black lines stand for $(g_{a\gamma\gamma}/m_{a})^2=1.406\times10^{-19}\,{\rm GeV}^{-2}\,{\rm eV}^{-2}$ for $E/N=23/6$ (case-I), $5.950\times10^{-20}\,{\rm GeV}^{-2}\,{\rm eV}^{-2}$ for $E/N=19/6$ (case-II), and $3.342\times10^{-22}\,{\rm GeV}^{-2}\,{\rm eV}^{-2}$ for $E/N=11/6$ (case-III), respectively. See more various supersymmetric and non-supersymmetric type models varying the parameter $E/N$ in Refs.\,\cite{Ahn:2015pia, Ahn:2016hbn}.}
\end{figure}

As seen in superpotential\,(\ref{lagrangian2}) since the SM charged-lepton fields (which are nontrivially $X$-charged Dirac fermions) have $U(1)_{\rm EM}$ charges, the axion $A_2$ coupling to electrons are added to the Lagrangian through a chiral rotation. And the axion $A_2$ couples directly to electrons, thereby the axion can be emitted by Compton scattering, atomic axio-recombination and axio-deexcitation, and axio-bremsstrahlung in electron-ion or electron-electron collisions\,\cite{Redondo:2013wwa}. The axion $A_2$ coupling to electron in the model reads
 \begin{eqnarray}
 g_{Aee}&=& \frac{|X_e|m_e}{f_{a_2}}\,, \quad\text{with}\,|X_e|=6
 \label{axion-electron0}
 \end{eqnarray}
 where $m_e=0.511$ MeV. Such weakly coupled flavored-axion $A_2$ has a wealth of interesting phenomenological implications in the context of astrophysics\,\footnote{From the cooling of white-dwarfs with the fine-structure of axion to electron, which is recently improved $4.1\times10^{-28}\lesssim\alpha_{Aee}\lesssim3.7\times10^{-27}$ in Ref.\,\cite{Bertolami:2014wua}, implying axion decay constant $f_{a_2}=(1.42-4.27)\times10^{10}\,{\rm GeV}$ and its corresponding QCD axion decay constant $F_A=(0.34-1.01)\times10^{10}\,{\rm GeV}$. See also the most recent analysis $\alpha_{Aee}=2.04^{+0.81}_{-0.77}\times10^{-27}$ at $1\sigma$\,\cite{ringward} leading to $f_{a2}=1.92^{+0.52}_{-0.29}\times10^{10}\,{\rm GeV}$ which is interpreted as $F_A=4.52^{+1.22}_{-0.69}\times10^{9}\,{\rm GeV}$. These hints including Ref.\,\cite{Isern:2008nt} seem incompatible with the bound in Eq.\,(\ref{cons_2}) from the decay process $K^+\rightarrow \pi+A_i$. However, if one relinquishes $N_{\rm DW}=1$ by considering $N_{\rm DW}>1$ in the case that the PQ phase transition happened during (or before) inflation, one can easily construct a model for accommodating the debating constraints under the present flavored-PQ scenario.}, like the formation of a cosmic diffuse background of axions from core collapse supernova explosions\,\cite{Raffelt:2011ft} or neutron star cooling\,\cite{Keller:2012yr}. 
There are several restrictive astrophysical limits\,\cite{PDG} on the axion models that couples to electrons, which arise from the above mentioned processes: among them, (i) from stars in the red giant branch of the color-magnitude diagram of globular clusters\,\cite{Redondo:2013wwa}, $\alpha_{Aee}<1.5\times10^{-26}$ ($95\%$ CL)\,\cite{Viaux:2013lha}, (ii) from white dwarfs (WDs) where bremsstrahlung is mainly efficient\,\cite{Raffelt:1985nj},  $\alpha_{Aee}<6\times10^{-27}$\,\cite{Bertolami:2014wua}, (iii) from the Sun the XENON100 experiment provides the upper bound, $g_{Aee}<7.7\times10^{-12}$ ($90\%$ CL)\,\cite{Aprile:2014eoa}, and recently (iv) from the solar flux the PandaX-II experiment provides the upper bound $g_{Aee}<4.35\times10^{-12}$ ($90\%$ CL)\,\cite{Fu:2017lfc}.
Here the fine-structure constant, $\alpha_{Aee}=g^{2}_{Aee}/4\pi$, is related to the axion-electron coupling constant $g_{Aee}$.
Then, the astrophysical lower bound of the PQ breaking scale $f_{a_2}$ and its corresponding QCD axion decay constant $F_A$ is derived from the above mentioned upper limits
\begin{eqnarray}
f_{a2}>(3.98\times10^8-1.23\times10^{10})\,{\rm GeV}\,\Leftrightarrow\,F_A>(9.38\times10^7-2.90\times10^{9})\,{\rm GeV}\,.
  \label{cons_1}
\end{eqnarray}
Since this limit for the QCD axion decay constant is much lower than the bound from $K^+\rightarrow\pi^++A_i$ in Eq.\,(\ref{k_bound}), we could not fix the scale of PQ phase transition.
Nevertheless, assuming that in the near future the NA62 experiment\,\cite{Fantechi:2014hqa} probes the flavored-axions,
 from the present upper bound in Eq.\,(\ref{k_bound}) and the future expected sensitivity of ${\rm Br}(K^+\rightarrow\pi^++A_i)$ in Eq.\,(\ref{knew_bound}) we can take the central value:% of $2.72\times10^{10}<F_A/[{\rm GeV}]<2.32\times10^{11}$:}
\begin{eqnarray}
 F_A=1.29\times10^{11}\,{\rm GeV}\,.
   \label{kk_new}
\end{eqnarray}
Hence, as shown in the left plot in FIG.\,\ref{Fig1}, the model for $F_A=1.29\times10^{11}$ GeV expected from the future sensitivity of ${\rm Br}(K^+\rightarrow\pi^+A_i)$ has predictions (horizontal solid-red, dashed-blue, and dotted-black lines crossed by solid-red (case-I), dotted-blue (case-II), and dashed-black (case-III) lines, respectively) on the QCD axion mass $m_a$ in terms of the Weinberg value $z=0.56$, and the pion decay constant $f_\pi=92$ MeV and $\mu m_u=(108.3\,{\rm MeV})^2z$, 
 \begin{eqnarray}
 m_a&=& \frac{f_\pi}{F_A}\Big(\frac{\mu m_u}{1+z+w}\Big)^{\frac{1}{2}}=45.8\,\mu{\rm eV}\,;
   \label{}
\end{eqnarray}
 its axion photon coupling expressed in terms of the axion mass, pion mass, pion decay constant, $z$ and $w$, 
 \begin{eqnarray}
 |g_{a\gamma\gamma}|&=&\frac{\alpha_{\rm em}}{2\pi}\frac{m_a}{f_\pi\,m_{\pi^0}}\frac{1}{\sqrt{F(z,w)}}\Big|\frac{E}{N}-\frac{2}{3}\frac{4+z+w}{1+z+w}\Big|\nonumber\\
 &=&\left\{\begin{array}{ll}
              1.72\times10^{-14}\,{\rm GeV}^{-1};& \text{~case-I}: ~( E/N=+23/6)  \\
              1.12\times10^{-14}\,{\rm GeV}^{-1};& \text{\,case-II}: \,(E/N=+19/6)\\
              8.37\times10^{-16}\,{\rm GeV}^{-1};& \text{case-III}: (E/N=+11/6)
             \end{array}\right\}\,.
   \label{}
\end{eqnarray}
The axion coupling to photon $g_{a\gamma\gamma}$ divided by the axion mass $m_{a}$ is dependent on $E/N$. Left plot in Fig.\,\ref{Fig1} shows the $E/N$ dependence of $(g_{a\gamma\gamma}/m_{a})^2$ so that the experimental limit is independent of the axion mass $m_{a}$\,\cite{Ahn:2014gva}: the values of $(g_{a\gamma\gamma}/m_{a})^2$ of our model are located lower than that of the experimentally excluded  bound $(g_{a\gamma\gamma}/m_{a})^2\leq1.44\times10^{-19}\,{\rm GeV}^{-2}\,{\rm eV}^{-2}$ from ADMX\,\cite{Asztalos:2003px}. For the Weinberg value $z=0.56$, the solid-red, dashed-blue, and dotted-black lines stand for $(g_{a\gamma\gamma}/m_{a})^2=1.406\times10^{-19}\,{\rm GeV}^{-2}\,{\rm eV}^{-2}$ for the anomaly value $E/N=23/6$ (case-I), $5.950\times10^{-20}\,{\rm GeV}^{-2}\,{\rm eV}^{-2}$ for $E/N=19/6$ (case-II), and $3.342\times10^{-22}\,{\rm GeV}^{-2}\,{\rm eV}^{-2}$ for $E/N=11/6$ (case-III), respectively.
%%%%%%%%%%%%%%%
%   Fig A-1   %
%%%%%%%%%%%%%%%
 %%%%%%%%%%%%%%%%%%%%%%%%%%%%%%%%%%%%%%%%%%%%%%%%%%%%%%%%%%%%%%%%%%%%%%%%%%%%
\subsection{Neutrinos}
 Even in the present model the quantum numbers $Q_{y^\nu_i}$ (or equivalently $Q_{L_{e,\mu,\tau}}$) are not uniquely determined through the model setup, together with the conditions above Eq.\,(\ref{cond_pseudo}) their quantum numbers can be assigned by their corresponding physical observables which are the pseudo-Dirac mass splittings $\Delta m^2_k$ responsible for new oscillations available on high-energy neutrinos through astronomical-scale baseline\,\cite{deGouvea:2009fp, Ahn:2016hbn, Anchordoqui:2004eb}.

As an explicit example, we take {\it case-I} in Eqs.\,(\ref{cond1}) and (\ref{caseI_lep}), and sequentially choose $y^\nu_i=\hat{y}^\nu_i\nabla^{9}_\Psi$ as 
 \begin{eqnarray}
     \begin{array}{ll}
             -Q_{y^\nu_1}=Q_{y^\nu_2}=Q_{y^\nu_3}=9q & \text{~~ for NO}  \\
             Q_{y^\nu_1}=-Q_{y^\nu_2}=Q_{y^\nu_3}=-9q & \text{~~~for IO}
             \end{array}\,,
 \label{}
 \end{eqnarray}
 by considering the conditions above Eq.\,(\ref{cond_pseudo}).
 At energies below the electroweak scale, all leptons obtain masses.
For the {\it case-I} the relevant lepton interaction terms with chiral fermions is given by 
 \begin{eqnarray}
  -{\cal L}^{\ell\nu}_{Y} &=&
   \overline{\ell_{R}}\,{\cal M}_{\ell}\,\ell_{L}+\frac{g}{\sqrt{2}}W^-_\mu\overline{\ell_{L}}\gamma^\mu\,\nu_{L}\nonumber\\
 &+& \frac{1}{2} \begin{pmatrix} \overline{\nu^c_L} & \overline{S_R} & \overline{N_R} \end{pmatrix} \begin{pmatrix} 0 & m^T_{DS} & m^T_D  \\  m_{DS} & e^{i\frac{A_{1}}{v_{\cal F}}}\, M_{S} & 0  \\ m_D & 0 & e^{i\frac{A_{1}}{v_{\cal F}}}\,M_R \end{pmatrix} \begin{pmatrix} \nu_L \\ S^c_R \\ N^c_R \end{pmatrix} +\text{h.c.}\,.
  \label{AxionLag2}
 \end{eqnarray}
And in the above Lagrangian\,(\ref{AxionLag2}) the charged-lepton and heavy Majorana neutrino mass terms read
 \begin{eqnarray}
  {\cal M}_{\ell}&=& {\left(\begin{array}{ccc}
 \hat{y}_{e}\nabla^6_{\Psi}\,e^{6i\frac{A_{2}}{v_{g}}} & 0 &  0 \\
 0 & \hat{y}_{\mu}\nabla^3_{\Psi}\,e^{-3i\frac{A_{2}}{v_{g}}} & 0 \\
 0 & 0 & \hat{y}_{\tau}\nabla_{\Psi}\,e^{i\frac{A_{2}}{v_{g}}}
 \end{array}\right)}v_{d}\,,\\
 \label{ChL3}
% \end{eqnarray}
% \begin{eqnarray}
 M_{R}&=&{\left(\begin{array}{ccc}
 1+\frac{2}{3}\tilde{\kappa}\,e^{i\phi} &  -\frac{1}{3}\tilde{\kappa}\,e^{i\phi} &  -\frac{1}{3}\tilde{\kappa}\,e^{i\phi} \\
 -\frac{1}{3}\tilde{\kappa}\,e^{i\phi} &  \frac{2}{3}\tilde{\kappa}\,e^{i\phi} &  1-\frac{1}{3}\tilde{\kappa}\,e^{i\phi}\\
 -\frac{1}{3}\tilde{\kappa}\,e^{i\phi} &  1-\frac{1}{3}\tilde{\kappa}\,e^{i\phi} &  \frac{2}{3}\tilde{\kappa}\,e^{i\phi}
 \end{array}\right)}M~,
 \label{MR1}
 \end{eqnarray}
 where
 \begin{eqnarray}
 \tilde{\kappa}\equiv\kappa\left|\frac{\hat{y}_{R}}{\hat{y}_\Theta}\right|,\quad\phi\equiv\arg\left(\frac{\hat{y}_{R}}{\hat{y}_{\Theta}}\right)~\,\text{with}~M\equiv \left|\hat{y}_{\Theta}\,\frac{v_{\Theta}}{\sqrt{2}}\right|.
 \label{MR2}
 \end{eqnarray}
 For NO, the Dirac and Majorana mass terms read
 \begin{eqnarray}
m_{DS}&=&{\left(\begin{array}{ccc}
 \hat{y}^s_{1}\,\nabla^{63}_{\Psi}\,e^{-63i\frac{A_{2}}{v_{g}}} &  0 &  0 \\
 0 &  \hat{y}^s_{2}\,\nabla^{18}_{\Psi}\,e^{18i\frac{A_{2}}{v_{g}}} &  0   \\
 0 &  0  &  \hat{y}^s_{3}\,\nabla^{17}_{\Psi}\,e^{17i\frac{A_{2}}{v_{g}}}
 \end{array}\right)}\,v_{u}, 
 \label{YDS1}\\
% \end{eqnarray}
%  \begin{eqnarray}
 M_{S}&=&{\left(\begin{array}{ccc}
 \hat{y}^{ss}_{1}\,\nabla^{144}_{\Psi}\,e^{-144i\frac{A_{2}}{v_{g}}} &  0 &  0 \\
 0 &  0 &  \hat{y}^{ss}_{2}\,\nabla^{54}_{\Psi}\,e^{54i\frac{A_{2}}{v_{g}}} \\
 0 &  \hat{y}^{ss}_{2}\,\nabla^{54}_{\Psi}\,e^{54i\frac{A_{2}}{v_{g}}} &  0 \end{array}\right)}\,\nabla_{\Theta}\,\frac{v_\Theta}{\sqrt{2}}\,,\label{YS1}\\
% \end{eqnarray}
% \begin{eqnarray}
 m_{D}%&=&{\left(\begin{array}{ccc}
 %\hat{y}^{\nu}_{1}\,e^{9i\frac{A_{2}}{v_{g}}} &  0 &  0 \\
 %0 &  0 &  \hat{y}^{\nu}_{2}\,e^{9i\frac{A_{2}}{v_{g}}}   \\
 %0 &  \hat{y}^{\nu}_{3}\,e^{-9i\frac{A_{2}}{v_{g}}}  &  0
 %\end{array}\right)}\nabla_T\,\nabla^{10}_{\Psi}\,v_{u}
 &=&\hat{y}^{\nu}_{1}{\left(\begin{array}{ccc}
 e^{9i\frac{A_{2}}{v_{g}}} &  0 &  0 \\
 0 &  0 &  y_{2}\,e^{-9i\frac{A_{2}}{v_{g}}}   \\
 0 &  y_{3}\,e^{-9i\frac{A_{2}}{v_{g}}}  &  0
 \end{array}\right)}\nabla_T\,\nabla^{9}_{\Psi}\,v_{u}, \label{Ynu1}
 \end{eqnarray}
where $y_{2}\equiv\hat{y}^{\nu}_{2}/\hat{y}^{\nu}_{1}$ and $y_{3}\equiv\hat{y}^{\nu}_{3}/\hat{y}^{\nu}_{1}$.
% \begin{eqnarray}
% y_{2}\equiv\frac{\hat{y}^{\nu}_{2}}{\hat{y}^{\nu}_{1}},\quad y_{3}\equiv\frac{\hat{y}^{\nu}_{3}}{\hat{y}^{\nu}_{1}}\,.
% \label{MR20}
% \end{eqnarray}
For IO, the Dirac and Majorana mass terms read
 \begin{eqnarray}
m_{DS}&=&{\left(\begin{array}{ccc}
 \hat{y}^s_{1}\,\nabla^{17}_{\Psi}\,e^{-17i\frac{A_{2}}{v_{g}}} &  0 &  0 \\
 0 &  \hat{y}^s_{2}\,\nabla^{17}_{\Psi}\,e^{17i\frac{A_{2}}{v_{g}}} &  0   \\
 0 &  0  &  \hat{y}^s_{3}\,\nabla^{28}_{\Psi}\,e^{-28i\frac{A_{2}}{v_{g}}}
 \end{array}\right)}\,v_{u}, 
 \label{YDS11}\\
% \end{eqnarray}
%  \begin{eqnarray}
 M_{S}&=&{\left(\begin{array}{ccc}
 \hat{y}^{ss}_{1}\,\nabla^{52}_{\Psi}\,e^{-52i\frac{A_{2}}{v_{g}}} &  0 &  0 \\
 0 &  0 &  \hat{y}^{ss}_{2}\,\nabla^{29}_{\Psi}\,e^{-29i\frac{A_{2}}{v_{g}}} \\
 0 &  \hat{y}^{ss}_{2}\,\nabla^{29}_{\Psi}\,e^{-29i\frac{A_{2}}{v_{g}}} &  0 \end{array}\right)}\,\nabla_{\Theta}\,\frac{v_\Theta}{\sqrt{2}}\,,\label{YS11}\\
% \end{eqnarray}
% \begin{eqnarray}
 m_{D}%&=&{\left(\begin{array}{ccc}
 %\hat{y}^{\nu}_{1}\,e^{9i\frac{A_{2}}{v_{g}}} &  0 &  0 \\
 %0 &  0 &  \hat{y}^{\nu}_{2}\,e^{9i\frac{A_{2}}{v_{g}}}   \\
 %0 &  \hat{y}^{\nu}_{3}\,e^{-9i\frac{A_{2}}{v_{g}}}  &  0
 %\end{array}\right)}\nabla_T\,\nabla^{10}_{\Psi}\,v_{u}
 &=&\hat{y}^{\nu}_{1}{\left(\begin{array}{ccc}
 e^{9i\frac{A_{2}}{v_{g}}} &  0 &  0 \\
 0 &  0 &  y_{2}\,e^{-9i\frac{A_{2}}{v_{g}}}   \\
 0 &  y_{3}\,e^{9i\frac{A_{2}}{v_{g}}}  &  0
 \end{array}\right)}\nabla_T\,\nabla^{9}_{\Psi}\,v_{u}\,. \label{Ynu11}
 \end{eqnarray}
Reminding that the hat Yukawa couplings in Eqs.\,(\ref{ChL3}-\ref{Ynu11}) are all of order unity and complex numbers. From Eq.\,(\ref{AxionLag2}), by redefining the light neutrino field $\nu_L$ as $P_\nu\,\nu_L$ and transforming $\ell_L\rightarrow P_\nu\,\ell_L$, $\ell_R\rightarrow P_\nu\,\ell_R$, $S_R\rightarrow P_s\,S_R$ where $P_{\nu, s}$ are diagonalized matrices of arbitrary phases, one can always make the Dirac neutrino Yukawa couplings $\hat{y}^\nu_1$, $y_2$, $y_3$ and $\hat{y}^s_1$, $\hat{y}^s_2$, $\hat{y}^s_3$ real and positive; then the parameters $\tilde{\kappa}$ and $y_{2,3}$ lie in the real and positive ranges
\begin{eqnarray}
0.17\lesssim\tilde{\kappa}\lesssim16.63\,,\qquad 0.1\lesssim y_{2,3}\lesssim 10\,,
  \label{tkr}
\end{eqnarray}
which will be used in numerical analysis, later.

After seesawing\,\cite{Ahn:2016hbn} due to the scale in Eq.\,(\ref{kk_new}) (or see Eqs.\,(\ref{k_bound}) and (\ref{cons_1})) much larger than the electroweak scale, in a basis where charged lepton and heavy neutrino masses are real and diagonal, we obtain
an effective light neutrino mass matrix in the basis $(\nu_L, S^c_R)$
\begin{eqnarray}
  {\cal M}_{\nu} ={\left(\begin{array}{cc}
  \delta_\nu  &  m^T_{\nu}   \\
  m_{\nu} &  M_S
  \end{array}\right)}.
  \label{eff_nu}
\end{eqnarray}
Under the given quantum numbers the active neutrinos appear as pseudo-Dirac neutrinos. And the pseudo-Dirac mass splittings in $k$-th pair $\Delta m^2_k\equiv m^{2}_{\nu_k}-m^2_{S_k}$ are expressed as
\begin{eqnarray}
  \Delta m^2_k=2\,m_k\,|\delta^\nu_k|\ll m_{\nu_k}
  \label{pseudo-D}
\end{eqnarray}
for all $k=1,2,3$, where $m_{\nu_k}$ and $m_{S_k}$ are mass eigenvalues of the effective mass matrix in Eq.\,(\ref{eff_nu}) and $\delta^\nu_k$ are mass eigenvalues of the seesaw formula $\delta_\nu=-m^{T}_{D}M^{-1}_{R}m_{D}$. Eq.\,(\ref{pseudo-D}) shows that
both the active neutrino masses $m_{\nu_k}$ coming from the matrix $m_\nu\equiv m_{DS}$ in Eq.\,(\ref{YDS1}) and the PMNS leptonic mixing angles coming from the matrix $\delta_\nu$ are closely tied to $\Delta m^2_k$ responsible for long wavelengths. 
Here the active neutrino masses we consider are given in Eq.\,(\ref{nu_mass_cn1}) for NO and Eq\,(\ref{nu_mass_ci1}) for IO.
On the other hand, the neutrino mixing parameters are determined by 
 \begin{eqnarray}
  \delta_{\nu} &=& -m^{T}_{D}M^{-1}_{R}m_{D}
  = m_{0}
   \small{\left(\begin{array}{ccc}
   1+2F & (1-F)y_{2} & (1-F)y_{3} \\
   (1-F)y_{2} & (1+\frac{F-3G}{2})y^{2}_{2} & (1+\frac{F+3G}{2})y_{2}y_{3}  \\
   (1-F)y_{3} & (1+\frac{F+3G}{2})y_{2}y_{3} & (1+\frac{F-3G}{2})y^2_{3}
   \end{array}\right)}\nonumber\\
   &=& U^\ast_{\rm PMNS}\,{\rm diag}(\delta^\nu_{1},\delta^\nu_{2},\delta^\nu_{3})\, U^{\dag}_{\rm PMNS}\,,
  \label{meff}
 \end{eqnarray}
 where the leptonic PMNS matrix $U_{\rm PMNS}$\,\cite{PDG} is given by Eq.\,(\ref{PMNS}), and
 \begin{eqnarray}
  F=(\tilde{\kappa}e^{i\phi}+1)^{-1}\,,\qquad G=(\tilde{\kappa}e^{i\phi}-1)^{-1}\,,\qquad m_0=\Big|\frac{\hat{y}^{\nu2}_1v^2_u}{3M}\Big|\nabla^2_T\nabla^{18}_\Psi\,.
  \label{nu_para}
 \end{eqnarray}
In the limit of $y_{2}, y_{3}\rightarrow1$ the above mass matrix reflects exact TBM mixing\,\cite{TBM} and its corresponding mass eigenvalues $|\delta^\nu_1|=3m_0|F|$, $|\delta^\nu_2|=3m_0$, $|\delta^\nu_3|=3m_0|G|$. Since in general it is expected deviations of $y_{2,3}$ from unity, Eq.\,(\ref{meff}) directly indicates that there could be deviations from the exact TBM, leading to a possibility to search for CP violation in neutrino oscillation experiments.
In addition, due to the small value of $\theta_{13}$ it is expected $|\delta^\nu_1|\simeq|\delta^\nu_2|\simeq|\delta^\nu_3|\approx3\,m_0$. To obtain the pseudo-Dirac mass splittings, taking the scale of heavy neutrino $M=\hat{y}_{\Theta}\,f_{a_1}/(|X_1|\sqrt{2(1+\kappa^2)})$ in Eq.\,(\ref{MR2})
 \begin{eqnarray}
  M\simeq2\times10^{11}\,{\rm GeV}
  \label{scale_seesaw}
 \end{eqnarray}
 from the QCD axion decay constant in Eq.\,(\ref{kk_new}) and using the best-fit values of the low energy neutrino oscillations\,\cite{Esteban:2016qun}, we can obtain the pseudo-Dirac mass splittings in a good approximation:
 \begin{eqnarray}
  \Delta m^2_3\simeq4.1\times10^{-14}\,{\rm eV}^2\,,\quad \Delta m^2_2\simeq7.1\times10^{-15}\,{\rm eV}^2\,,\quad \Delta m^2_1\simeq3.5\times10^{-36}\,{\rm eV}^2\,,
  \label{Spn}
 \end{eqnarray}
for NO with $\hat{y}^s_1=1$;
 \begin{eqnarray}
  \Delta m^2_2\simeq4.1\times10^{-14}\,{\rm eV}^2\simeq \Delta m^2_1\,,\qquad \Delta m^2_3\simeq2.5\times10^{-22}\,{\rm eV}^2\,,
  \label{Spi}
 \end{eqnarray}
for IO with $\hat{y}^s_3=1$.

Due to the precise measurement of $\theta_{13}$, which is relatively large, it may now be possible to put constraints on the Dirac phase $\delta_{CP}$ which will be obtained in the long baseline neutrino oscillation experiments T2K\,\cite{t2k}, NO$\nu$A\,\cite{nova}, MINOS\,\cite{minos} etc.. However, the current large uncertainty on $\theta_{23}$ is at present limiting the information that can be extracted from the $\nu_{e}$ appearance measurements. Precise measurements of all the mixing angles, especially $\theta_{23}$, are needed to maximize the sensitivity to the leptonic CP violation. The active neutrino oscillation experiments are now on a new step to confirm the CP violation and octant of atmospheric mixing angle $\theta_{23}$ in the lepton sector. Actually, the recent data of T2K and NO$\nu$A experiments indicate a finite CP phase\,\cite{finite_CP} together with their preferred octant on $\theta_{23}$\,\cite{t2k, nova}. 

%%%%%%%%%%%%%%%%%%%%%%%%%%%%%%%%%%%%%%%%%%%%%%%%%%%%%%%%%%%%%%%%%%%%%%
\subsection{Numerical analysis for neutrino mixing parameters}
\label{num}
In order to show model predictions on the leptonic Dirac CP phase $\delta_{CP}$ incident to the atmospheric mixing angle $\theta_{23}$, we perform a  numerical simulation by using the linear algebra tools of Ref.\,\cite{Antusch:2005gp} with the $3\sigma$ constraints of the low energy neutrino oscillations\,\cite{Esteban:2016qun}.% with the value $\nabla_\Psi$ obtained in quark sector, see Eq.\,(\ref{quarkvalue}). 

In our numerical analysis, we take\,\footnote{From Eqs.\,(\ref{kk_new}) and (\ref{MR2}) we reasonably well square the axion decay constant $f_{a_1}$ with the scale $M$.} $M=2\times10^{11}$ GeV in Eq.\,(\ref{scale_seesaw}) and $\tan\beta=7.40$ (see Eq.\,(\ref{kk_new}) and Eq.\,(\ref{tanpara})), as inputs. %Recalling that all the hat Yukawa couplings are of order unity, {\it i.e.}, $1/\sqrt{10}\lesssim|\hat{y}|\lesssim\sqrt{10}$. 
The seesaw formula in Eq.\,(\ref{meff}) for obtaining neutrino mixing parameters ($\theta_{12}$, $\theta_{13}$, $\theta_{23}$, $\delta_{CP}$) and their eigenvalues $\delta^\nu_k=\Delta m^{2}_k/2m_{\nu_k}$ ($k=1,2,3$) contains seven parameters: $y_{1}(\equiv\hat{y}^{\nu}_{1}\nabla_T\,\nabla_\Psi^{9}),v_{u}, \,M,\,y_{2},\,y_{3},\,\tilde{\kappa},\,\phi$. The first three ($y_{1}$, $M,$ and $v_{u}$) lead to the overall scale parameter $m_{0}$ in Eq.\,(\ref{nu_para}), which is closely related to the $U(1)_{X_1}$ breaking scale. The next four ($y_2,\,y_3,\,\tilde{\kappa},\,\phi$) with the allowed ranges in Eq.\,(\ref{tkr}) give rise to the deviations from TBM, the CP phases, and corrections to the pseudo-Dirac mass splittings $\Delta m^{2}_{k}=2m_k\,|\delta_k|$. Since the individual neutrino masses ($m_{\nu_k}=m_k$) are determined as in Eqs.\,(\ref{nu_mass_cn1}) and (\ref{nu_mass_ci1}) within the $3\sigma$ constraints of the low energy neutrino oscillations\,\cite{Esteban:2016qun}, for numerical simulation we can simply fix the pseudo-Dirac mass splittings\,\footnote{They may be fixed by high energy astronomical-baseline experiments, such as IceCube\,\cite{icecube}.} $\Delta m^{2}_{k}=2m_k\,|\delta_k|$, without loss of generality, as in Eq.\,(\ref{Spn}) for NO and Eq.\,(\ref{Spi}) for IO. Then, the active neutrino masses $m_{\nu_k}$ can directly be linked to the eigenvalues $\delta^\nu_{k}$ in Eq.\,(\ref{meff}).

Hence, there are only left the five physical parameters $m_{0},y_{2},y_{3},\tilde{\kappa},\phi$ contained in Eq.\,(\ref{meff}), which can be determined from the $3\sigma$ experimental bounds of three mixing angles ($\theta_{12},\theta_{13},\theta_{23}$) and two active neutrino mass splittings ($\Delta m^2_{\rm Sol}$, $\Delta m^2_{\rm Atm}$). 
Among nine observables (six mixing parameters $\theta_{12}, \theta_{23}, \theta_{13}, \delta_{CP}, \varphi_{1,2}$ and three mass eigenvalues $m_{\nu_1}, m_{\nu_2}, m_{\nu_3}$) in low energy neutrino sector, the remaining four observables (one Dirac CP phase $\delta_{CP}$, two Majorana CP phases $\varphi_{1,2}$, and one active neutrino mass) can be predicted in the model. Here both the lightest active neutrino mass and the Majorana CP phases contributing to the effective active neutrino masses are negligibly small enough in the model.
Therefore, we can have reasonable model predictions on the Dirac CP phase $\delta_{CP}$ incident to behavior of the large uncertainty on $\theta_{23}$.

The recent analysis based on global fits\,\cite{Gonzalez-Garcia:2015qrr, Esteban:2016qun, Gariazzo:2018pei} of the neutrino oscillations enters into a new phase of precise determination of mixing angles and mass squared differences: we take the global fits at $3\sigma$\,\cite{Esteban:2016qun}, shown in Table\,\ref{exp}, as experimental constraints.
\begin{table}[h]
%\begin{widetext}
%\begin{center}
\caption{\label{exp} The global fit of three-flavor oscillation parameters at $3\sigma$ level\,\cite{Esteban:2016qun}. NO = normal neutrino mass ordering; IO = inverted mass ordering. And $\Delta m^{2}_{\rm Sol}\equiv m^{2}_{\nu_2}-m^{2}_{\nu_1}$, $\Delta m^{2}_{\rm Atm}\equiv m^{2}_{\nu_3}-m^{2}_{\nu_1}$ for NO, and  $\Delta m^{2}_{\rm Atm}\equiv m^{2}_{\nu_2}-m^{2}_{\nu_3}$ for IO.}
\begin{ruledtabular}
\begin{tabular}{cccccccccccc} &$\theta_{13}[^{\circ}]$&$\delta_{CP}[^{\circ}]$&$\theta_{12}[^{\circ}]$&$\theta_{23}[^{\circ}]$&$\Delta m^{2}_{\rm Sol}[10^{-5}{\rm eV}^{2}]$&$\Delta m^{2}_{\rm Atm}[10^{-3}{\rm eV}^{2}]$\\
\hline
$3\,\sigma$$\begin{array}{ll}
\hbox{NO}\\
\hbox{IO}
\end{array}$&$\begin{array}{ll}
7.99\rightarrow8.90 \\
8.03\rightarrow8.93
\end{array}$&$0\rightarrow360$&~$31.38\rightarrow35.99$&$\begin{array}{ll}
38.4\rightarrow52.8 \\
38.8\rightarrow53.1
\end{array}$
 &$7.03\rightarrow8.09$&$ \begin{array}{ll}
                           2.407\rightarrow2.643 \\
                           2.399\rightarrow2.635
                          \end{array}$\\
\end{tabular}
\end{ruledtabular}
%\end{center}
%\end{widetext}
\end{table}

\begin{figure}[t]
%\vspace*{-5.0cm}
%\hspace*{-1cm}
\begin{minipage}[h]{7.3cm}
\epsfig{figure=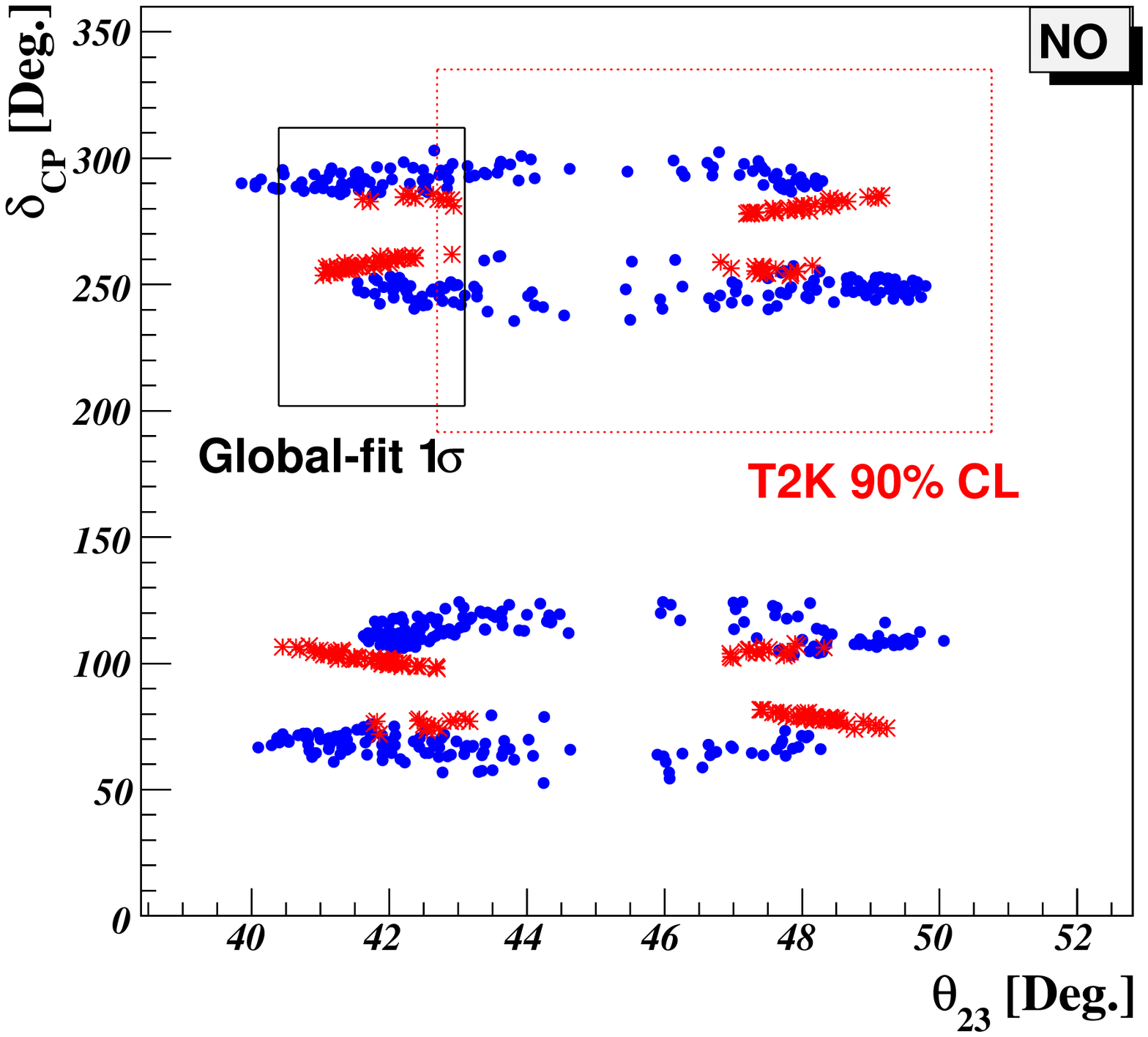,width=7.3cm,angle=0}
\end{minipage}
\hspace*{1.0cm}
\begin{minipage}[h]{7.3cm}
\epsfig{figure=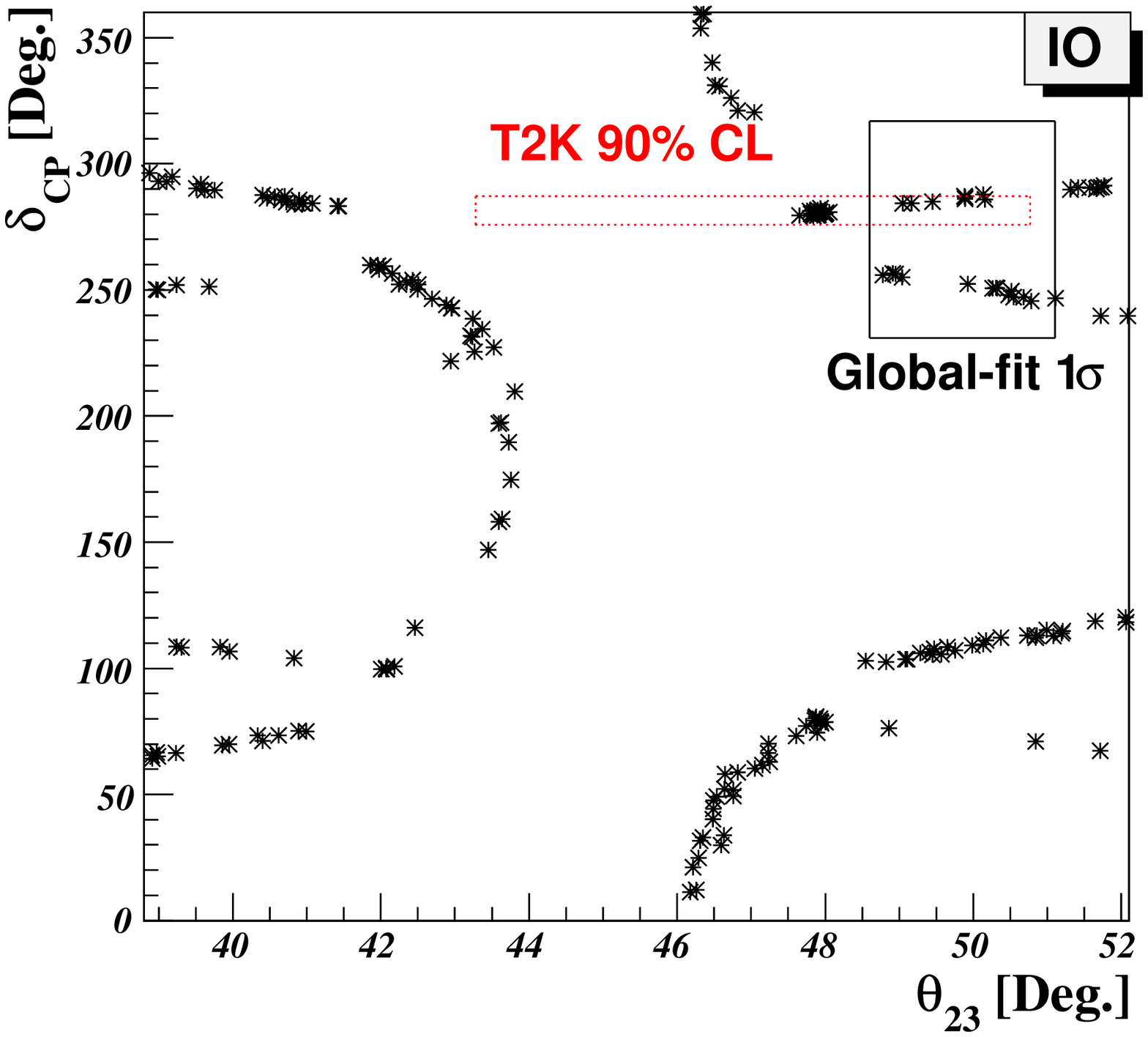,width=7.3cm,angle=0}
\end{minipage}
%\vspace*{-5.5cm}
\caption{\label{Fig2} Plot for leptonic Dirac CP phase $\delta_{CP}$ as a function of the atmospheric mixing angle $\theta_{23}$. In the left plot for NO, red-asters indicate the case of $\Delta m^2_2/\Delta m^2_3\geq\sqrt{\Delta m^{2}_{\rm Sol}/\Delta m^{2}_{\rm Atm}}$ and blue-spots for $\Delta m^2_2/\Delta m^2_3<\sqrt{\Delta m^{2}_{\rm Sol}/\Delta m^{2}_{\rm Atm}}$. In the left plot, black-quadrangle represents global-fit $1\sigma$ bounds $\delta_{CP}/[^\circ]=261^{+51}_{-59}$ and $\theta_{23}/[^\circ]=41.6^{+1.5}_{-1.2}$\,\cite{Esteban:2016qun}, while red-dotted quadrangle favored by T2K\,\cite{Abe:2017vif} stands for 90\% CL bounds $\delta_{CP}/[^\circ]=[191.0, 334.8]$ and $\theta_{23}/[^\circ]=47.9^{+2.9}_{-5.2}$. Right plot for IO,  where black-quadrangle for global-fit $1\sigma$ bounds $\delta_{CP}/[^\circ]=277^{+40}_{-46}$ and $\theta_{23}/[^\circ]=50.0^{+1.1}_{-1.4}$\,\cite{Esteban:2016qun}, while red-dotted quadrangle favored by T2K\,\cite{Abe:2017vif} for 90\% CL bounds $\delta_{CP}/[^\circ]=[275.8, 287.2]$ and $\theta_{23}/[^\circ]=47.9^{+2.9}_{-4.6}$.}
\end{figure}

Scanning all the parameter spaces ($0.17\lesssim\tilde{\kappa}\lesssim16.63$, $0.1\lesssim y_{2,3}\lesssim 10$
  in Eq.\,(\ref{tkr}), $1/\sqrt{10}\lesssim\hat{y}^\nu_1\lesssim\sqrt{10}$, and $0\leq\phi\leq2\pi$) by putting the experimental $3\sigma$ constraints in Table\,\ref{exp} with the above input parameters:\\
\noindent For NO with the setting of pseudo-Dirac mass splittings $\Delta m^2_3=4.1\times10^{-14}\,{\rm eV}^2$, $\Delta m^2_2=7.1\times10^{-15}\,{\rm eV}^2$, $\Delta m^2_1=3.5\times10^{-36}\,{\rm eV}^2$ in Eq.\,(\ref{Spn}) the neutrino parameter spaces are fixed as
 \begin{align}
  &\tilde{\kappa} \in [0.17,0.36] ,
  & \phi \in [91^{\circ},95^{\circ}]\cup[265^{\circ},270^{\circ}]\,,
  \nonumber\\
  & \hat{y}^\nu_{1} \in [1.06,1.15],
  & y_{2} \in [0.87,1.12],\qquad\qquad\quad
  & y_{3} \in [0.89,1.12]\,,
  \label{input1}
 \end{align}
 for $\Delta m^2_2/\Delta m^2_3\geq m_{\nu_2}/m_{\nu_3}$ (or equivalently $\delta^{\nu}_2/\delta^{\nu}_3\geq1$), indicating red-asters in the left plot of FIG.\,\ref{Fig2};
  \begin{align}
  &\tilde{\kappa} \in [0.17,0.30] ,
  & \phi \in [85^{\circ},100^{\circ}]\cup[265^{\circ},274^{\circ}]\,,
  \nonumber\\
  & \hat{y}^\nu_{1} \in [1.06,1.10],
  & y_{2} \in [0.93,1.12],\qquad\qquad\quad
  & y_{3} \in [0.22,1.12]\,,
  \label{input01}
 \end{align}
 for $\Delta m^2_2/\Delta m^2_3<m_{\nu_2}/m_{\nu_3}$ (or equivalently $\delta^{\nu}_2/\delta^{\nu}_3<1$), indicating blue-spots in the left plot of FIG.\,\ref{Fig2}.\\
\noindent  For IO with the setting of pseudo-Dirac mass splittings $\Delta m^2_2=4.1\times10^{-14}\,{\rm eV}^2=\Delta m^2_1$, $\Delta m^2_3=2.5\times10^{-22}\,{\rm eV}^2$ in Eq.\,(\ref{Spi}) we obtain
 \begin{align}
  &\tilde{\kappa} \in [0.17,0.66] ,
  & \phi \in [92^{\circ},110^{\circ}]\cup[260^{\circ},268^{\circ}]\,,
  \nonumber\\
  & \hat{y}^\nu_{1} \in [1.06,1.13],
  & y_{2} \in [0.80,1.20],\qquad\qquad\quad
  & y_{3} \in [0.81,1.21]\,.
  \label{input2}
 \end{align}
As shown in FIG.\,\ref{Fig2} there are remarkable predictions on $\delta_{CP}$ as a function of the atmospheric mixing angle $\theta_{23}$ for NO (left plot) and IO (right plot). Moreover, in the model, the neutrinoless-double-beta ($0\nu\beta\beta$)-decay rate effectively measures the absolute value of the $ee$-component of the effective neutrino mass matrix ${\cal M}_\nu$ in Eq.\,(\ref{eff_nu}) in the basis where the charged lepton mass matrix is real and diagonal, which can be expressed as $|m_{ee}|=|\sum^3_{k=1}(U_{ek}/\sqrt{2})^2(m_{\nu_k}-m_{S_k})|$. Thus, accurate measurements of $\theta_{23}$ and $\delta_{CP}$ are crucial for a test of our model.
In addition, the discovery of $0\nu\beta\beta$-decay in the on-going or future $0\nu\beta\beta$-decay experiments\,\cite{nuBB}, with sensitivities $0.01<|m_{ee}|/{\rm eV}<0.1$, will rule out the present model.

%%%%%%%%%%%%%%%
%   Fig A-2   %
%%%%%%%%%%%%%%%
In the left plot (NO) of FIG.\,\ref{Fig2}, with the sum of neutrino masses $\sum^3_{i=1}m_{\nu_i}\ni[0.058, 0.060]\,\text{eV}$ and the amplitude of $0\nu\beta\beta$-decay rate $|m_{ee}|\simeq4\times10^{-13}$ eV, the red-asters stand for predictions on 
 \begin{align}
  &\theta_{23}\ni[40.5^\circ, 43.2^\circ]\cup[48.8^\circ, 49.2^\circ], \nonumber\\
  & \delta_{CP}\ni[72.1^\circ, 81.7^\circ]\cup[98.0^\circ, 107.8^\circ]\cup[253.6^\circ, 262.0^\circ]\cup[278.0^\circ, 285.8^\circ]\,,
  \label{result1}
 \end{align}
 for $\Delta m^2_2/\Delta m^2_3\geq m_{\nu_2}/m_{\nu_3}$; similarly, the blue-spots indicate predictions on 
 \begin{align}
  &\theta_{23}\ni[40.0^\circ, 44.2^\circ]\cup[45.4^\circ, 50.1^\circ], \nonumber\\
  & \delta_{CP}\ni[52.5^\circ, 79.5^\circ]\cup[103.3^\circ, 124.4^\circ]\cup[236.0^\circ, 257.4^\circ]\cup[285.4^\circ, 303.0^\circ]\,,
  \label{result2}
 \end{align}
 for $\Delta m^2_2/\Delta m^2_3< m_{\nu_2}/m_{\nu_3}$. On the other hand, in the right plot (IO) of FIG.\,\ref{Fig2} with the sum of neutrino masses $\sum^3_{i=1}m_{\nu_i}\ni(0.097, 0.102)\,\text{eV}$ and the amplitude of $0\nu\beta\beta$-decay rate $|m_{ee}|\simeq4\times10^{-13}$ eV, the black-crosses stand for predictions on 
 \begin{align}
  &\delta_{CP}\ni[65.0^\circ, 295.5^\circ]\,,\quad\text{for}~
   \theta_{23}\ni[38.9^\circ, 43.8^\circ]\,;\\
  &\delta_{CP}\ni[11.3^\circ, 120.0^\circ]\cup[235.7^\circ, 360.0^\circ]\,, \quad\text{for}~
   \theta_{23}\ni[46.2^\circ, 52.8^\circ]\,.
  \label{result3}
 \end{align}
Even the results for IO look like having wide ranges, as shown in the right plot (IO) of FIG.\,\ref{Fig2} there is a remarkable predictive-pattern for $\delta_{CP}$ as a function of $\theta_{23}$.

 %%%%%%%%%%%%%%%%%%%%%%%%%%%%%%%%%%%%%%%%%%%%%%%%%%%%%%%%%%%%%%%%%%%%%%%%%%%%
\section{Conclusion}
We have constructed a minimalistic SUSY model for quarks, leptons, and flavored-axions (and its one linear combination, QCD axion) through the argument that the $U(1)$ mixed-gravitational anomaly cancellation could be of central importance in constraining the fermion content of a new chiral gauge theory. It contains a flavor-structured $G_F=SL_2(F_3)\times U(1)_X$ symmetry for a compact description of new physics beyond SM.
We have showed that axionic domain-wall condition $N_{\rm DW}$ with the $U(1)_X$ mixed-gravitational anomaly cancellation depends on both $U(1)_X$ charged quark and lepton flavors; the scale of PQ symmetry breakdown congruent to the seesaw scale is constrained through constraints coming from astrophysics and particle physics.  
Along this line, we have showed that the model could well be flavor-structured by the $G_F$ symmetry in a unique way that domain-wall number $N_{\rm DW}=1$ with the $U(1)_X$ mixed-gravitational anomaly-free condition demands additional Majorana fermions as well as the flavor puzzles of SM are well delineated by new expansion parameters defined by the model dependent parameters, $U(1)_X$ charges and $U(1)_X$-$[SU(3)_C]^2$ anomaly coefficients. In turn, we have showed that the flavored-axion model works well by performing a numerical simulation for the quark sector, leading to $\tan\beta=7.40$ with the experimental results of the CKM mixing angles and their corresponding quark masses satisfied, as shown in Sec.\,\ref{num_quark}. 

And we have showed that the constraint on the $U(1)_X$ symmetry breaking scale coming from the particle physics on the rare decay $K^+\rightarrow \pi^++A_i$ is much stronger than that from the astroparticle physics on QCD axion and flavored-axion cooling of stars.
So, in order to fix the scale of PQ phase transition we take a testable QCD axion decay constant, $F_A=1.29\times10^{11}$ GeV, from the current bound and the future expected sensitivity on ${\rm Br}(K^+\rightarrow\pi^++A_i)$, which gives model predictions on the axion mass $m_a=45.8\,\mu$eV and axion-photon couplings $|g_{a\gamma\gamma}|=1.72\times10^{-14}\,{\rm GeV}^{-1}$ for $E/N=+23/6$ (case-I), 1$.12\times10^{-14}\,{\rm GeV}^{-1}$ for $E/N=+19/6$ (case-II), and $8.37\times10^{-16}\,{\rm GeV}^{-1}$ for $E/N=+11/6$ (case-III), as summarized in FIG.\,\ref{Fig1} for QCD axion. 

Subsequently, we have showed that the lepton sector structured by the symmetry $G_F$ provides 
 interesting physical implications on neutrino: hierarchical mass spectra and unmeasurable neutrinoless-double-beta decay rate with the interesting predictions on atmospheric mixing angle and leptonic Dirac CP phase favored by the recent long-baseline neutrino oscillation experiments, as summarized in FIG.\,\ref{Fig2} for NO and IO.

\newpage
%%%%%%%%%%%%%%%%%%%%%%%%%%%%%%%%%%%%%%%%%%%%%%%%%%%%%%%%%%%%%%%%%%%%%%%%%
\appendix
%%%%%%%%%%%%%%%%%%%%%%%%%%%%%%%%%%%%%%%%%%%%%%
\section{The $SL_2(F_3)$ group}
\label{SL2F3}
The $SL_2(F_3)$ is the double covering of the tetrahedral group $A_4$\,\cite{slf3, aranda, Feruglio:2007uu}. It contains 24 elements and has three kinds of representations: one triplet {\bf 3} and three singlets ${\bf 1}$, ${\bf 1}'$ and ${\bf 1}''$, and three doublets ${\bf 2}$, 
${\bf 2}'$ and ${\bf 2}''$. The representations ${\bf 1}'$, ${\bf 1}''$ and ${\bf 2}'$, ${\bf 2}''$ are complex conjugated to each
other. Note that $A_4$ is not a subgroup of $SL_2(F_3)$, since the two-dimensional representations
cannot be decomposed into representations of $A_4$. The generators $S$ and $T$ satisfy the required conditions $S^2=R$, $T^3=1$, $(ST)^3=1$, and $R^2=1$, where $R =1$ in case of the odd-dimensional representation and $R=-1$ for ${\bf 2}$, 
${\bf 2}'$ and ${\bf 2}''$ such that $R$ commutes with all elements of the group. The matrices $S$ and $T$ representing the generators depend on the representations of the group\,\cite{aranda}:
                    \begin{equation}
                        \begin{array}{ccccc}
                            1 &\qquad& S=1 &\quad& T=1\\
                            1^\prime &\qquad& S=1 &\quad& T=\omega\\
                            1^{\prime\prime} &\qquad& S=1 &\quad& T=\omega^2\\[10pt]
                            2 &\qquad& S=A_1 &\quad& T=\omega A_2\\
                            2^\prime &\qquad& S=A_1 &\quad& T=\omega^2 A_2\\
                            2^{\prime\prime} &\qquad& S=A_1 &\quad& T=A_2\\[10pt]
                            3 &\qquad& S=\dfrac{1}{3}\left(\begin{array}{ccc}
                                                            -1 & 2\omega & 2\omega^2 \\
                                                            2\omega^2 & -1 & 2\omega \\
                                                            2\omega & 2\omega^2 & -1 \\
                                                            \end{array}\right)
                                    &\quad& T=\left(\begin{array}{ccc}
                                                            1 & 0 & 0 \\
                                                            0 & \omega & 0 \\
                                                            0 & 0 & \omega^2 \\
                                                        \end{array}\right)
                        \end{array}\nonumber
                    \end{equation}
        where we have used the matrices
            \begin{equation}
            A_1=-\dfrac{1}{\sqrt{3}}\left(\begin{array}{cc}
                                                i & \sqrt2e^{i\pi/12} \\
                                                -\sqrt2e^{-i\pi/12} & -i \\
                                            \end{array}\right)\,\qquad
            A_2=\left(\begin{array}{cc}
                          \omega & 0 \\
                          0 & 1 \\
                        \end{array}\right)\;.\nonumber
            \end{equation}\\
        \indent The following multiplication rules between the various representations are calculated in Ref.\,\cite{aranda}, where
         $\alpha_{i}$ indicate the elements of the first representation of the product and $\beta_{i}$ indicate those of the second representation. Moreover $a,b=0,\pm1$ and we denote $1^0\equiv1$, $1^1\equiv1^\prime$, $1^{-1}\equiv1^{\prime\prime}$ and similarly for the doublet representations. On the right-hand side the sum $a+b$ is modulo 3.\\
        \indent  The multiplication rules with the 1-dimensional representations are the following:
            \[
                \begin{array}{l}
                1\otimes Rep=Rep\otimes1=Rep\quad\text{with $Rep$ whatever
                representation}\\[8pt]
                1^a\otimes1^b=1^b\otimes1^a=1^{a+b}\equiv\alpha\beta\\[8pt]
                1^a\otimes2^b=2^b\otimes1^a=2^{a+b}\equiv\left(\begin{array}{cc}
                                                                                \alpha\beta_1, &
                                                                                \alpha\beta_2 \\
                                                                        \end{array}\right)\\[-10pt]\\[8pt]

                1^\prime\otimes3=3=\left(\begin{array}{ccc}
                                            \alpha\beta_3, &
                                            \alpha\beta_1, &
                                            \alpha\beta_2\\
                                    \end{array}\right)\,,
                \qquad1^{\prime\prime}\otimes3=3=\left(\begin{array}{ccc}
                                            \alpha\beta_2, &
                                            \alpha\beta_3, &
                                            \alpha\beta_1\\
                                    \end{array}\right)\,.
                \end{array}
            \]
        \indent  The multiplication rules with the 2-dimensional representations are
            \[
                \begin{array}{lc}
                &2\otimes2=2^\prime\otimes2^{\prime\prime}=2^{\prime\prime}\otimes2^\prime=3\oplus1\\
                \text{with}\quad
                                  &  3=\left(\begin{array}{ccc}
                                        \dfrac{1-i}{2}(\alpha_1\beta_2+\alpha_2\beta_1), &
                                        i\alpha_1\beta_1, &
                                        \alpha_2\beta_2
                                    \end{array}\right)\,,\qquad
                                    1=\alpha_1\beta_2-\alpha_2\beta_1\,;
                \\[-8pt]\\
                &2\otimes2^\prime=2^{\prime\prime}\otimes2^{\prime\prime}=3\oplus1^\prime\\
                \text{with}\quad
                                   & 3=\left(\begin{array}{ccc}
                                        \alpha_2\beta_2,&
                                        \dfrac{1-i}{2}(\alpha_1\beta_2+\alpha_2\beta_1),&
                                        i\alpha_1\beta_1
                                    \end{array}\right)\,,\qquad
                                    1^\prime=\alpha_1\beta_2-\alpha_2\beta_1\,;
                \\[-8pt]\\
                &2\otimes2^{\prime\prime}=2^\prime\otimes2^\prime=3\oplus1^{\prime\prime}\\
                \text{with}\quad
                                   & 3=\left(\begin{array}{ccc}
                                        i\alpha_1\beta_1, &
                                        \alpha_2\beta_2, &
                                        \dfrac{1-i}{2}(\alpha_1\beta_2+\alpha_2\beta_1)
                                    \end{array}\right)\,,\qquad
                                    1^{\prime\prime}=\alpha_1\beta_2-\alpha_2\beta_1\,;
                \\[-8pt]\\
                 \end{array}
            \]               
                 \[
                \begin{array}{lc}
                &2\otimes3=2\oplus2^\prime\oplus2^{\prime\prime}\\
                \text{with}
                                    &2\;=\;\left(\begin{array}{cc}
                                                (1+i)\alpha_2\beta_2+\alpha_1\beta_1, &
                                                (1-i)\alpha_1\beta_3-\alpha_2\beta_1
                                            \end{array}\right)\\
                                   & 2^\prime=\;\left(\begin{array}{cc}
                                                (1+i)\alpha_2\beta_3+\alpha_1\beta_2, &
                                                (1-i)\alpha_1\beta_1-\alpha_2\beta_2
                                            \end{array}\right)\\
                                   &\;\; 2^{\prime\prime}=\left(\begin{array}{cc}
                                            (1+i)\alpha_2\beta_1+\alpha_1\beta_3, &
                                            (1-i)\alpha_1\beta_2-\alpha_2\beta_3
                                             \end{array}\right)\,;
                \\[-8pt]\\
                &2^\prime\otimes3=2\oplus2^\prime\oplus2^{\prime\prime}\\
                \text{with}\quad
                                   & 2\;\,=\left(\begin{array}{cc}
                                                (1+i)\alpha_2\beta_1+\alpha_1\beta_3, &
                                                (1-i)\alpha_1\beta_2-\alpha_2\beta_3
                                            \end{array}\right)\\
                                    &2^\prime\,=\left(\begin{array}{cc}
                                                (1+i)\alpha_2\beta_2+\alpha_1\beta_1, &
                                                (1-i)\alpha_1\beta_3-\alpha_2\beta_1
                                            \end{array}\right)\\
                                    &\;\;2^{\prime\prime}=\left(\begin{array}{cc}
                                                (1+i)\alpha_2\beta_3+\alpha_1\beta_2, &
                                                (1-i)\alpha_1\beta_1-\alpha_2\beta_2
                                            \end{array}\right)\,;
                \\[-8pt]\\
                                 \end{array}
            \]               
                 \[
                \begin{array}{lc}
               & 2^{\prime\prime}\otimes3=2\oplus2^\prime\oplus2^{\prime\prime}\\
                \text{with}\quad
                                    &2\;\,=\left(\begin{array}{cc}
                                                    (1+i)\alpha_2\beta_3+\alpha_1\beta_2, &
                                                    (1-i)\alpha_1\beta_1-\alpha_2\beta_2
                                                \end{array}\right)\\
                                    &2^\prime\,=\left(\begin{array}{cc}
                                                    (1+i)\alpha_2\beta_1+\alpha_1\beta_3, &
                                                    (1-i)\alpha_1\beta_2-\alpha_2\beta_3
                                                \end{array}\right)\\
                                    &\;\;2^{\prime\prime}=\left(\begin{array}{cc}
                                                    (1+i)\alpha_2\beta_2+\alpha_1\beta_1, &
                                                    (1-i)\alpha_1\beta_3-\alpha_2\beta_1
                                                \end{array}\right)\,.
                \end{array}
            \]\\

        \indent   The multiplication rule with the 3-dimensional representations is
            \[
                3\otimes3=3_S\oplus3_A\oplus1\oplus1^\prime\oplus1^{\prime\prime}
            \]
        where
            \begin{gather*}
                3_S=\frac{1}{3}\left(\begin{array}{ccc}
                                         2\alpha_1\beta_1-\alpha_2\beta_3-\alpha_3\beta_2, &
                                         2\alpha_3\beta_3-\alpha_1\beta_2-\alpha_2\beta_1, &
                                         2\alpha_2\beta_2-\alpha_1\beta_3-\alpha_3\beta_1\\
                                \end{array}\right)\,\\
                3_A=\frac{1}{2}\left(\begin{array}{ccc}
                                         \alpha_2\beta_3-\alpha_3\beta_2, &
                                         \alpha_1\beta_2-\alpha_2\beta_1, &
                                         \alpha_3\beta_1-\alpha_1\beta_3\\
                                \end{array}\right)\\[10pt]
                \begin{array}{l}
                1\;\,=\alpha_1\beta_1+\alpha_2\beta_3+\alpha_3\beta_2\\
                1^\prime\,=\alpha_3\beta_3+\alpha_1\beta_2+\alpha_2\beta_1\\
                1^{\prime\prime}=\alpha_2\beta_2+\alpha_1\beta_3+\alpha_3\beta_1\;.
                \end{array}
            \end{gather*}
%%%%%%%%%%%%%%%%%%%%%%%%%%%%%%%%%%%%%%%%%%%%%%
\section{Higher order corrections}
 \label{corre}
We consider possible next-to-leading order corrections. Higher-dimensional operators invariant under $SL_2(F_3)\times U(1)_X$ symmetry, suppressed by additional powers of the cutoff scale $\Lambda$, could be added to the leading order terms in the superpotential. Then the mass and mixing matrices for fermions can be corrected by both a shift of the vacuum configuration and nontrivial next-to-leading operators contributing to the Yukawa superpotential.  

For example, we show that next leading corrections to the renormalizable Majorana neutrino sector can well be under control.
 In addition to the leading order Yukawa superpotential $W_{\ell\nu}$, we should also consider those higher dimensional operators that could be induced by the flavon fields $\Phi_{T}$ and $\eta$ which are not charged under the $U(1)_{X}$.
At the next leading order in the Majorana neutrino sector those operators triggered by the field $\Phi_{T}$ are written as $(N^cN^c\Theta\Phi_{T})_{{\bf 1}}/\Lambda$ and $(N^cN^c\Phi_{S}\Phi_{T})_{{\bf 1}}/\Lambda$.
Here the first term, after symmetry breaking, is absorbed into the leading order terms in the renormalizable superpotential and the corresponding Yukawa couplings are redefined. On the other hand, the second term could be non-trivial and it can be clearly expressed as
\begin{eqnarray}
\Delta W_{\nu}&=&\frac{\hat{y}^{R}_{1}}{2\Lambda}(N^cN^c)_{{\bf 1}}(\Phi_{S}\Phi_{T})_{{\bf 1}}+\frac{\hat{y}^{R}_{2}}{2\Lambda}(N^cN^c)_{{\bf 1}'}(\Phi_{S}\Phi_{T})_{{\bf 1}''}+\frac{\hat{y}^{R}_{3}}{2\Lambda}(N^cN^c)_{{\bf 1}''}(\Phi_{S}\Phi_{T})_{{\bf 1}'}\nonumber\\
&+&\frac{\hat{y}^{R}_{s}}{2\Lambda}(N^cN^c)_{{\bf 3}_s}(\Phi_{S}\Phi_{T})_{{\bf 3}_s}+\frac{\hat{y}^{R}_{a}}{2\Lambda}(N^cN^c)_{{\bf 3}_s}(\Phi_{S}\Phi_{T})_{{\bf 3}_a}\,.
\label{NewMR}
\end{eqnarray}
Indeed at order $1/\Lambda$, after symmetry breaking, there is a new structure contributing to $M_{R}$, whose contribution is written as
 \begin{eqnarray}
 \Delta M_{R}&=&\nabla_T\,{\left(\begin{array}{ccc}
 \tilde{\kappa}_{1}+\frac{4}{9}\tilde{\kappa}_{s} & \tilde{\kappa}_{2}+\frac{1}{9}\tilde{\kappa}_{s}-\frac{1}{6}\tilde{\kappa}_{a} &  \tilde{\kappa}_{3}+\frac{1}{9}\tilde{\kappa}_{s}+\frac{1}{6}\tilde{\kappa}_{a} \\
 \tilde{\kappa}_{2}+\frac{1}{9}\tilde{\kappa}_{s}-\frac{1}{6}\tilde{\kappa}_{a} &  \tilde{\kappa}_{3}-\frac{2}{9}\tilde{\kappa}_{s}-\frac{1}{3}\tilde{\kappa}_{a} &  \tilde{\kappa}_{1}-\frac{2}{9}\tilde{\kappa}_{s}\\
 \tilde{\kappa}_{3}+\frac{1}{9}\tilde{\kappa}_{s}+\frac{1}{6}\tilde{\kappa}_{a} &  \tilde{\kappa}_{1}-\frac{2}{9}\tilde{\kappa}_{s} &  \tilde{\kappa}_{2}-\frac{2}{9}\tilde{\kappa}_{s}+\frac{1}{3}\tilde{\kappa}_{a}
 \end{array}\right)}M~,
 \label{MRCorr}
 \end{eqnarray}
where 
 \begin{eqnarray}
 \tilde{\kappa}_{i}\equiv\kappa\,\hat{y}^R_i/\hat{y}_{\Theta}
 \end{eqnarray}
 with $i=1,2,3,s,a$. Even though these corrections to the leading order picture seem to non-trivial, these can be kept small, below few percent level due to $\nabla_T$ in Eq.\,(\ref{quarkvalue}) by keeping $|\hat{y}_R|\gtrsim|\hat{y}^R_i|$, {\it i.e.} $\tilde{\kappa}\gtrsim\tilde{\kappa}_i$ with Eq.\,(\ref{MR2}). Then, eventually, after seesawing in Eq.\,(\ref{meff}) the active neutrino mixing matrix at leading order could not be crucially changed.

Next, considering higher dimensional operators induced by $\Phi_{T},\Phi_{S},\Theta, \Psi,\eta$ invariant under $SL_{2}(F_3)\times U(1)_X$ in the driving superpotential $W_{v}$, which are suppressed by additional powers of the cut-off scale $\Lambda$, they can lead to small deviations from the leading order vacuum configurations.
The next leading order superpotential $\delta W_{v}$, which is linear in the driving fields and invariant under $SL_{2}(F_3)\times U(1)_{X}\times U(1)_R$, is given by
 \begin{eqnarray}
 \delta W_{v}&=& \frac{1}{\Lambda}\Big\{a_{1}(\Phi_{T}\Phi_{T})_{{\bf 3s}}(\Phi_{T}\Phi^{T}_{0})_{{\bf 3a}}+a_{2}(\Phi_{T}\Phi_{T})_{{\bf 1}}(\Phi_{T}\Phi^{T}_{0})_{{\bf 1}}+a_{3}(\Phi_{T}\Phi_{T})_{{\bf 1}'}(\Phi_{T}\Phi^{T}_{0})_{{\bf 1}''}\nonumber\\
 &+&a_{4}(\Phi_{T}\Phi_{T})_{{\bf 1}''}(\Phi_{T}\Phi^{T}_{0})_{{\bf 1}'}+a_{5}\Psi\tilde{\Psi}(\Phi_{T}\Phi^{T}_{0})_{{\bf 1}}+a_{6}(\eta\Phi_{T})_{{\bf 2}}(\eta\Phi^{T}_0)_{{\bf 2}}+a_{7}(\eta\Phi_{T})_{{\bf 2}'}(\eta\Phi^{T}_0)_{{\bf 2}''}\Big\}\nonumber\\
 &+&\frac{1}{\Lambda}\Big\{b_{1}(\Phi_{S}\Phi_{S})_{{\bf 3s}}(\Phi_{T}\Phi^{S}_{0})_{{\bf 3a}}
 +b_{2}(\Phi_{S}\Phi_{S})_{{\bf 3s}}(\Phi_{T}\Phi^{S}_{0})_{{\bf 3s}}+b_{3}(\Phi_{S}\Phi_{S})_{{\bf 1}}(\Phi_{T}\Phi^{S}_{0})_{{\bf 1}}\nonumber\\
 &+&b_{4}(\Phi_{S}\Phi_{S})_{{\bf 1}'}(\Phi_{T}\Phi^{S}_{0})_{{\bf 1}''}+b_{5}(\Phi_{S}\Phi_{S})_{{\bf 1}''}(\Phi_{T}\Phi^{S}_{0})_{{\bf 1}'}+b_{6}\Phi^{S}_{0}(\Phi_{S}\Phi_{T})_{{\bf 3a}}\Theta\nonumber\\
 &+&b_{7}\Phi^{S}_{0}(\Phi_{S}\Phi_{T})_{{\bf 3s}}\Theta+b_{8}\Phi^{S}_{0}(\Phi_{S}\Phi_{T})_{{\bf 3a}}\tilde{\Theta}+b_{9}\Phi^{S}_{0}(\Phi_{S}\Phi_{T})_{{\bf 3s}}\tilde{\Theta}
 \nonumber\\
 &+&b_{10}(\Phi^{S}_{0}\Phi_{T})_{{\bf 1}}\Theta\Theta
 +b_{11}(\Phi^{S}_{0}\Phi_{T})_{{\bf 1}}\Theta\tilde{\Theta}+b_{12}(\Phi^{S}_{0}\Phi_{T})_{{\bf 1}}\tilde{\Theta}\tilde{\Theta} \Big\}\nonumber\\
 &+&\frac{\Theta_{0}}{\Lambda}\left\{c_{1}(\Phi_{S}\Phi_{S})_{{\bf 3s}}\Phi_{T}+c_{2}(\Phi_{S}\Phi_{T})_{{\bf 1}}\tilde{\Theta}\right\}+\frac{\Psi_{0}}{\Lambda}d_{1}(\Phi_{T}\Phi_{T})_{{\bf 3s}}\Phi_{T}\nonumber\\
 &+& \frac{1}{\Lambda}\Big\{f_{1}(\eta\eta)_{{\bf 3}}(\eta\eta_{0})_{{\bf 3}}+f_{2}(\Phi_{T}\Phi_{T})_{{\bf 3s}}(\eta\eta_{0})_{{\bf 3}}
 +f_{3}(\Phi_{T}\Phi_{T})_{{\bf 1}}(\eta\eta_{0})_{{\bf 1}}
 +%f_{4}(\Phi_{S}\Phi_{S})_{{\bf 1}}(\eta\eta_{0})_{{\bf 1}}+
 f_{4}\Psi\tilde{\Psi}(\eta\eta_{0})_{{\bf 1}}\Big\}\,.
 \label{Npotential}
 \end{eqnarray}
By keeping only the first order in the expansion, one can obtain the minimization equations.
The corrections to the VEVs, Eqs.\,(\ref{vevdirection1},\ref{vevdirection2},\ref{vevdirection3}), are of relative order $1/\Lambda$ and affect the flavon fields $\Phi_{S}$,  $\Phi_{T}$, $\Theta$, $\tilde{\Theta}$, $\eta$ and $\Psi$, and the vacuum configuration can be modified with relations among the dimensionless parameters ($a_{1}...a_{7}$, $b_{1}...b_{12}$, $c_{1}, c_{2}$, $d_{1}$, $f_{1}...f_{4}$).
Given the ranges for $\nabla_Q$ with $Q=\eta,S,T,\Theta,\Psi$ in Eq.\,(\ref{quarkvalue}), one can expect that the shifts $|\delta\tilde{\Theta}|, |\delta\Theta|/v_{\Theta},|\delta v_{S_i}|/v_S,|\delta v_{T_i}|/v_T, |\delta v_{\eta_i}|/v_\eta, |\delta v_{\Psi}|/v_\Psi$. can be kept small enough, below a few percent level. Then the mixing angles of the active neutrinos in Eq.\,(\ref{meff}) may not be crucially modified by the next-to-leading order results in FIG.\,\ref{Fig2} for NO and IO.

%%%%%%%%%%%%%%%%%%%%%%%%%%%%%%%%%%%%%%%%%%%%%%
\section{The leptonic mixing matrix}
 \label{}
In the mass eigenstate basis the PMNS leptonic mixing matrix\,\cite{PDG} at low energies is visualized in the charged weak interaction, which is expressed in terms of three mixing angles, $\theta_{12}, \theta_{13}, \theta_{23}$, and three \cp-odd phases (one $\delta_{CP}$ for the Dirac neutrino and two $\varphi_{1,2}$ for the Majorana neutrino) as
 \begin{eqnarray}
  U_{\rm PMNS}=
  {\left(\begin{array}{ccc}
   c_{13}c_{12} & c_{13}s_{12} & s_{13}e^{-i\delta_{CP}} \\
   -c_{23}s_{12}-s_{23}c_{12}s_{13}e^{i\delta_{CP}} & c_{23}c_{12}-s_{23}s_{12}s_{13}e^{i\delta_{CP}} & s_{23}c_{13}  \\
   s_{23}s_{12}-c_{23}c_{12}s_{13}e^{i\delta_{CP}} & -s_{23}c_{12}-c_{23}s_{12}s_{13}e^{i\delta_{CP}} & c_{23}c_{13}
   \end{array}\right)}P_{\nu}~,
 \label{PMNS}
 \end{eqnarray}
where $s_{ij}\equiv \sin\theta_{ij}$, $c_{ij}\equiv \cos\theta_{ij}$ and $P_{\nu}$ is a diagonal phase matrix what is that particles are Majorana ones.

%%%%%%%%%%%%%%%%%%%%%%%%%%%%%%%%%%%%%%%%%%%%%%
\section{Axionic domain-wall condition}
 \label{axi_do}
The quantum numbers associated to charged-leptons are assigned to enforce a positive value of ``electromagnetic anomaly ($U(1)_X$-$[U(1)_{\rm EM}]^2$)/color anomaly ($U(1)_X$-$[SU(3)_{C}]^2$) coefficient'' within the range\,\footnote{This range is derived from the bound ADMX experiment\,\cite{Asztalos:2003px} $(g_{a\gamma\gamma}/m_{a})^2\leq1.44\times10^{-19}\,{\rm GeV}^{-2}\,{\rm eV}^{-2}$.} $0<E/N<4$:
 \begin{eqnarray}
  \frac{E}{N}&=&\frac{23}{6}\,,\qquad\text{for}~{\cal Q}_{y_\tau}=-q, ~{\cal Q}_{y_\mu}=3q, ~{\cal Q}_{y_e}=-6q\,;~\text{case-I}\\
   \frac{E}{N}&=&\frac{19}{6}\,,\qquad\text{for}~{\cal Q}_{y_\tau}=q, \quad{\cal Q}_{y_\mu}=3q, ~{\cal Q}_{y_e}=-6q\,;~\text{case-II}\\
 \frac{E}{N}&=&\frac{11}{6}\,,\qquad\text{for}~{\cal Q}_{y_\tau}=-q, ~{\cal Q}_{y_\mu}=-3q, ~{\cal Q}_{y_e}=6q\,;~\text{case-III}
 \label{cas}
 \end{eqnarray}
where $E=\sum_f(\delta^{\rm G}_2X_{1f}+\delta^{\rm G}_1X_{2f})(Q^{\rm em}_{f})^2$ and $N=2\delta^{\rm G}_1\delta^{\rm G}_2$.
Then, in terms of ${\cal Q}_{\cal Y}$ the anomaly-free condition of $U(1)_X\times[gravity]^2$ is expressed as
 \begin{eqnarray}
 U(1)_X\times[gravity]^2&\propto&3\left\{4p-{\cal Q}_{y_b}+2({\cal Q}_{Y_s}-{\cal Q}_{Y_d}-{\cal Q}_{y_c}-{\cal Q}_{y_s})\right\}_{\rm quark}\nonumber\\
 &+&\left\{3p-{\cal Q}_{y^s_1}-{\cal Q}_{y^s_2}-{\cal Q}_{y^s_3}-{\cal Q}_{y_e}-{\cal Q}_{y_\mu}-{\cal Q}_{y_\tau}\right\}_{\rm lepton}=0\,.
 \label{ux_gr}
 \end{eqnarray}
 This vanishing anomaly, however, does not restrict ${\cal Q}_{y^{\nu}_i}$ (or equivalently ${\cal Q}_{y^{ss}_i}$), whose quantum  numbers can be constrained by the  new neutrino oscillations of astronomical-scale baseline, which will be shown later.
With the given above $U(1)_X$ quantum numbers, such $U(1)_X\times[gravity]^2$ anomaly is free for 
 \begin{eqnarray}
 15\,\frac{X_1}{2}=k_2\,X_2\qquad\text{with}~ k_2=\left\{\begin{array}{ll}
              \tilde{{\cal Q}}_{y^s_1}+\tilde{{\cal Q}}_{y^s_2}+\tilde{{\cal Q}}_{y^s_3}-13;& \text{case-I}  \\
              \tilde{{\cal Q}}_{y^s_1}+\tilde{{\cal Q}}_{y^s_2}+\tilde{{\cal Q}}_{y^s_3}-11;& \text{case-II}\\
              \tilde{{\cal Q}}_{y^s_1}+\tilde{{\cal Q}}_{y^s_2}+\tilde{{\cal Q}}_{y^s_3}-7;& \text{case-III}
             \end{array}\right\}\,.
\label{cond1}
 \end{eqnarray}
 where $\tilde{{\cal Q}}_{y^s_i}={\cal Q}_{y^s_1}/X_2$.  We take $k_2=\pm15$ for the $U(1)_{X_i}$ charges to be smallest
making no axionic domain-wall problem. Hence, for $\tilde{{\cal Q}}_{y^s_1}+\tilde{{\cal Q}}_{y^s_2}+\tilde{{\cal Q}}_{y^s_3}=28$ $(-2)$ for the case-I; 26 ($-4$) for the case-II; 22 ($-8$) for the case-III, the values of $k_i$ are rescaled as
\begin{eqnarray}
 k_1=\pm k_2=1\,,
 \label{k_val}
\end{eqnarray}
with $p=k_2$ and $q=k_1$ by $k_1\,p=k_2\,q=k_1\,k_2$.
In the present model the color anomaly coefficients are given by $\delta^{\rm G}_1=2X_1$ and $\delta^{\rm G}_2=3X_2$. 
Then, the axionic domain-wall condition in Eq.\,(\ref{dw}) is rewritten as
\begin{eqnarray}
 N_1=4\,,\quad N_2=3\,,
  \label{dw2}
\end{eqnarray}
ensuring that no axionic domain-wall problem occurs.

%%%%%%%%%%%%%%%%%%%%%%%%%%%%%%%%%%%%%%%%%%%%%%%%%%%%%%%%%%%%%%%%%%%%%%%%%%%%%%%%%%%%%%%%%%%%%%%%%%
\acknowledgments
{We thank prof. Hai-Yang Cheng and Xue Xun for useful discussions and kind hospitality, and MH Ahn for useful comments on axion part. This work is supported by the NSFC under Grant No. U1738209. 
}

%%%%%%%%%%%%%%%%%%%%%%%%%%%%%%%%%%%%%%%%%%%%%%%%%%%%%%%%%%%%%%%%%%%%%%%%%%%%%%%%%%%%%

\end{document}